\pdfoutput=0
\documentclass[11pt]{article}
\usepackage{axodraw}
\usepackage{epsfig}
\usepackage{amsfonts}
\usepackage{amsmath}
\usepackage{bm,bbm}
\usepackage{cite}
\renewcommand{\baselinestretch}{1.15} %{1.15} 
\allowdisplaybreaks[1]

\input pix.sty

\newcommand{\rephase}{\phi^{+}_{(a)\I\J}}        %% {\phi^{ }_{\I\J}}
\newcommand{\imphase}{\phi^{-}_{(a)\I\J}}  %% {\bar{\phi}^{ }_{\I\J}}

\renewcommand{\B}{\rmii{$B$}}
\newcommand{\I}{\rmii{$I$}}
\newcommand{\J}{\rmii{$J$}}
\newcommand{\sL}{\rmii{$L$}}
\newcommand{\K}{\rmii{$K$}}
\newcommand{\V}{\mathcal{V}}
\newcommand{\T}{\rmii{$T$}}
\newcommand{\muY}{\mu_\rmii{$Y$}}
\newcommand{\muA}{\mu_\rmii{\!$A$}}
\newcommand{\muB}{\mu_\rmii{$B$}}
\newcommand{\muQ}{\mu_\rmii{$Q$}}
\newcommand{\muZ}{\mu_\rmii{$Z$}}
\newcommand{\munLa}{\mu_{\nu_{\rmii{$L$}a}}}
\newcommand{\mueLa}{\mu_{e_{\rmii{$L$}a}}}
\newcommand{\muphin}{\mu_{\phi^{ }_{0}}}
\newcommand{\muphip}{\mu_{\phi^{ }_{+}}}
\newcommand{\muW}{\mu_\rmii{$W^+$}}
\newcommand{\muZn}{\mu_\rmii{$Z^0$}}

\newcommand{\aL}{a^{ }_\rmii{L}}
\newcommand{\aR}{a^{ }_\rmii{R}}
\renewcommand{\eq}{eq.~}
\renewcommand{\eqs}{eqs.~}
\renewcommand{\se}{sec.~}
\renewcommand{\ses}{secs.~}
\renewcommand{\fig}{fig.~}
\renewcommand{\figs}{figs.~}

\newcommand{\rmO}{{\mathcal{O}}}
\newcommand{\bmu}{\bar\mu}

\def\lsi{\raise0.3ex\hbox{$<$\kern-0.75em\raise-1.1ex\hbox{$\sim$}}}
\def\gsi{\raise0.3ex\hbox{$>$\kern-0.75em\raise-1.1ex\hbox{$\sim$}}}
\newcommand{\lsim}{\mathop{\lsi}}
\newcommand{\gsim}{\mathop{\gsi}}

\newcommand{\nF}{n_\rmii{F}}

\newcommand{\rmii}[1]{{\mbox{\tiny\rm{#1}}}}
\newcommand{\rmiii}[1]{{\mbox{\tiny{$\scriptstyle{\rm#1}$}}}}
\newcommand{\re}{\mathop{\mbox{Re}}}
\newcommand{\im}{\mathop{\mbox{Im}}}
\newcommand{\Tint}[1]{{\hbox{$\sum$}\!\!\!\!\!\!\!\int\,}_{\!\!\!\!\raise-0.9ex\hbox{$\scriptstyle{#1}$}}}
\newcommand{\Tinti}[1]{{{\Sigma}\!\!\!\!\raise0.3ex\hbox{$\int$}_\rmii{${#1}$}}}

\newcommand{\bi}{\begin{itemize}}
\newcommand{\ei}{\end{itemize}}
\newcommand{\hide}[1]{ }
\newcommand{\bsl}[1]{\,\slash\!\!\!\!{#1}\,}
\newcommand{\msl}[1]{\,\slash\!\!\!{#1}\,}
\newcommand{\scat}[1]{\mbox{scat}^{ }_\rmi{$#1$}}
\newcommand{\Sq}[2]{#1 - #2^2} %% {#2^2-#1} 
\newcommand{\s}[1]{s_{#1}}

\newcommand{\E}{\mathcal{E}}
\newcommand{\U}{\mathcal{U}}
\newcommand{\Proj}{\mathcal{I}}
\renewcommand{\P}{\mathcal{P}}
\renewcommand{\K}{\mathcal{K}}
\newcommand{\MM}{M^2}

\newcommand{\mS}{m_\phi} % {m_\rmiii{S}}
\newcommand{\mtop}{m_t}
\newcommand{\mbot}{m_b}

\newcommand{\mh}{m_h}
\newcommand{\mZ}{m_\rmii{$Z$}} % {m_\rmiii{Z}}
\newcommand{\mB}{m_\rmii{$B$}} % {m_\rmiii{B}}
\newcommand{\mW}{m_\rmii{$W$}} % {m_\rmiii{W}}
\newcommand{\mZp}{m_\rmii{$Z$'}} % {m_\rmiii{Z'}}
\newcommand{\mWp}{m_\rmii{$W$'}} % {m_\rmiii{W'}}
\newcommand{\mGZ}{m_\rmii{$GZ$}} % {m_\rmiii{GZ}}
\newcommand{\mGW}{m_\rmii{$GW$}} % {m_\rmiii{GW}}
\newcommand{\ana}{{\nu^{ }_a}}
\newcommand{\aea}{{e^{ }_a}}
\newcommand{\ala}{{\ell^{ }_a}}
\newcommand{\an}{{\nu^{ }_b}}
\newcommand{\aeb}{{e^{ }_b}}
\newcommand{\atopL}{{q_\rmii{L}^{ }}} 
\newcommand{\atopR}{{t_\rmii{R}^{ }}}

\newcommand{\aS}{\phi} % {{S}}
\renewcommand{\atop}{t}
\newcommand{\abot}{b}
\newcommand{\ado}{d}
\newcommand{\aup}{u}
\newcommand{\ah}{h}
\newcommand{\aZ}{Z}
\newcommand{\aQ}{\gamma} % {Q}

\newcommand{\aW}{W}
\newcommand{\Ana}{{\bar{\nu}^{ }_a}}
\newcommand{\Aea}{{\bar{e}^{ }_a}}
\newcommand{\An}{{\bar{\nu}^{ }_b}}
\newcommand{\Ae}{{\bar{e}^{ }_b}}
\newcommand{\AtopL}{{\bar{q}_\rmii{L}^{ }}} 
\newcommand{\AbotR}{{\bar{b}_\rmii{R}^{ }}}
\newcommand{\Atop}{\bar{t}}
\newcommand{\Abot}{\bar{b}}
\newcommand{\Ado}{\bar{d}}
\newcommand{\Aup}{\bar{u}}
\newcommand{\AW}{\bar{W}}
\newcommand{\htop}{h_t}
\newcommand{\hbot}{h_b}
\newcommand{\lam}{\lambda}
\newcommand{\gt}{\tilde{g}}
\newcommand{\cW}{c_\rmiii{W}}
\newcommand{\sW}{s_\rmiii{W}}
\newcommand{\cWW}{c_\rmiii{2W}}

\def\TAsc(#1,#2)(#3,#4,#5)%
{\SetWidth{2.0}\CArc(#1,#2)(#3,#4,#5)\SetWidth{1.0}}
\def\Lwidth{3}

\def\TAgl(#1,#2)(#3,#4,#5){\SetWidth{2.0}\PhotonArc(#1,#2)(#3,#4,#5){\Lwidth}%
{6.283 #3 mul 360 div #4 #5 sub #4 #5 sub mul sqrt mul Tdensity mul}%
\SetWidth{1.0}}
\def\TLgl(#1,#2)(#3,#4){\SetWidth{2.0}\Photon(#1,#2)(#3,#4){\Lwidth}
{#1 #3 sub #1 #3 sub mul #2 #4 sub #2 #4 sub mul add sqrt Tdensity mul}%
\SetWidth{1.0}}
\renewcommand{\picc}[1]{\;\parbox[c]{60pt}{\begin{picture}(60,30)(0,-10)
\SetWidth{1.0}\SetScale{1.3} #1 \end{picture}}\; }
\def\Lwidth{1.3}

\renewcommand{\pic}[1]{\;\parbox[c]{30pt}{\begin{picture}(30,30)(0,-3)
\SetWidth{1.0}\SetScale{0.8} #1 \end{picture}}\;}
\renewcommand{\picb}[1]{\;\parbox[c]{45pt}{\begin{picture}(45,30)(0,-3)
\SetWidth{1.0}\SetScale{0.8} #1 \end{picture}}\;}
\def\decAA{\picb{%
 \Line(-5,14)(5,14)%
 \Line(-5,16)(5,16)%
 \Line(3,13)(7,17)%
 \Line(3,17)(7,13)%
 \Line(5,15)(15,15)%
 \Lgl(15,15)(35,25)%
 \Lqu(15,15)(35,5)%
}}
\def\decBB{\picb{%
 \Line(5,14)(16,14)%
 \Line(5,16)(16,16)%
 \Lsc(16,16)(35,25)%
 \Lqu(16,14)(35,5)%
}}
\def\decA{\picb{%
 \Line(-5,14)(5,14)%
 \Line(-5,16)(5,16)%
 \Line(3,13)(7,17)%
 \Line(3,17)(7,13)%
 \Line(5,15)(15,15)%
 \Lqu(15,15)(45,30)%
 \Lgl(15,15)(30,7.5)%
 \Lqu(30,7.5)(45,0)%
 \Laqu(30,7.5)(45,15)%
}}
\def\decB{\picb{%
 \Line(5,14)(16,14)%
 \Line(5,16)(16,16)%
 \Lqu(16,16)(45,30)%
 \Lsc(16,14)(30,7.5)%
 \Laqu(30,7.5)(45,15)%
 \Lqu(30,7.5)(45,0)%
}}
\def\decC{\picb{%
 \Line(-5,14)(5,14)%
 \Line(-5,16)(5,16)%
 \Line(3,13)(7,17)%
 \Line(3,17)(7,13)%
 \Line(5,15)(15,15)%
 \Lqu(15,15)(45,0)%
 \Lgl(15,15)(30,22.5)%
 \Lqu(30,22.5)(45,30)%
 \Laqu(30,22.5)(45,15)%
}}
\def\decD{\picb{%
 \Line(-5,14)(5,14)%
 \Line(-5,16)(5,16)%
 \Line(3,13)(7,17)%
 \Line(3,17)(7,13)%
 \Line(5,15)(15,15)%
 \Lqu(15,15)(45,30)%
 \Lgl(15,15)(30,7.5)%
 \Lgl(30,7.5)(45,15)%
 \Lsc(30,7.5)(45,0)%
}}
\def\decE{\picb{%
 \Line(5,14)(16,14)%
 \Line(5,16)(16,16)%
 \Lqu(16,16)(45,30)%
 \Lsc(16,14)(30,7.5)%
 \Lgl(30,7.5)(45,15)%
 \Lsc(30,7.5)(45,0)%
}}
\def\decF{\picb{%
 \Line(5,14)(16,14)%
 \Line(5,16)(16,16)%
 \Line(16,16)(30,22.5)%
 \Lqu(30,22.5)(45,30)%
 \Lgl(30,22.5)(45,15)%
 \Lsc(16,14)(45,0)%
}}
\def\decG{\picb{%
 \Line(-5,14)(5,14)%
 \Line(-5,16)(5,16)%
 \Line(3,13)(7,17)%
 \Line(3,17)(7,13)%
 \Line(5,15)(15,15)%
 \Line(15,15)(30,22.5)%
 \Lqu(30,22.5)(45,30)%
 \Lgl(30,22.5)(45,15)%
 \Lgl(15,15)(45,0)%
}}
\def\decH{\picb{%
 \Line(-5,14)(5,14)%
 \Line(-5,16)(5,16)%
 \Line(3,13)(7,17)%
 \Line(3,17)(7,13)%
 \Line(5,15)(15,15)%
 \Line(15,15)(27,24)%
 \Lqu(27,24)(45,30)%
 \Lgl(15,15)(45,15)%
 \Lgl(27,24)(31,18.66)%
 \Lgl(36,12)(45,0)%
}}
\def\decI{\picb{%
 \Line(5,14)(16,14)%
 \Line(5,16)(16,16)%
 \Lqu(16,16)(45,30)%
 \Lsc(16,14)(30,7.5)%
 \Lgl(30,7.5)(45,15)%
 \Lgl(30,7.5)(45,0)%
}}
\def\decJ{\picb{%
 \Line(-5,14)(5,14)%
 \Line(-5,16)(5,16)%
 \Line(3,13)(7,17)%
 \Line(3,17)(7,13)%
 \Line(5,15)(15,15)%
 \Lqu(15,15)(45,30)%
 \Lgl(15,15)(30,7.5)%
 \Lgl(30,7.5)(45,0)%
 \Lgl(30,7.5)(45,15)%
}}
\def\decK{\picb{%
 \Line(5,14)(16,14)%
 \Line(5,16)(16,16)%
 \Lqu(16,16)(45,30)%
 \Lsc(16,14)(30,7.5)%
 \Lsc(30,7.5)(45,15)%
 \Lsc(30,7.5)(45,0)%
}}
\def\selfGmaster{\pic{%
 \Lgl(0,15)(13,15)%
 \Lgl(17,15)(30,15)%
 \CCirc(15,15){5}{Black}{Gray}
}}
\def\selfA{\pic{%
 \Lgl(0.5,15)(7.5,15)%
 \Lgl(22.5,15)(29.5,15)%
 \PhotonArc(15,15)(7,0,360){1.3}{11}
}}
\def\selfB{\pic{%
 \Lgl(0,15)(8,15)%
 \Lgl(22,15)(30,15)%
 \Agh(15,15)(7,0,180)
 \Agh(15,15)(7,180,360)
}}
\def\selfD{\pic{%
 \Lgl(0,15)(8,15)%
 \Lgl(22,15)(30,15)%
 \Asc(15,15)(7,0,180)
 \Asc(15,15)(7,180,360)
}}
\def\selfF{\pic{%
 \Lgl(0,15)(8,15)%
 \Lgl(22,15)(30,15)%
 \Asc(15,15)(7,0,180)
 \Agl(15,15)(7,180,360)
}}
\def\selfG{\pic{%
 \Lgl(0,15)(8,15)%
 \Lgl(22,15)(30,15)%
 \Aqu(15,15)(7,0,180)
 \Aqu(15,15)(7,180,360)
}}
\def\selfSmaster{\pic{%
 \Lsc(0,15)(13,15)%
 \Lsc(17,15)(30,15)%
 \CCirc(15,15){5}{Black}{Gray}
}}
\def\selfSA{\pic{%
 \Lsc(0,15)(8,15)%
 \Lsc(23,15)(30,15)%
 \PhotonArc(15,15)(7,0,360){1.3}{11}
}}
\def\selfSB{\pic{%
 \Lsc(0,15)(8,15)%
 \Lsc(22,15)(30,15)%
 \Agh(15,15)(7,0,180)
 \Agh(15,15)(7,180,360)
}}
\def\selfSD{\pic{%
 \Lsc(0,15)(8,15)%
 \Lsc(22,15)(30,15)%
 \Asc(15,15)(7,0,180)
 \Asc(15,15)(7,180,360)
}}
\def\selfSF{\pic{%
 \Lsc(0,15)(8,15)%
 \Lsc(22,15)(30,15)%
 \Asc(15,15)(7,0,180)
 \Agl(15,15)(7,180,360)
}}
\def\selfSG{\pic{%
 \Lsc(0,15)(8,15)%
 \Lsc(22,15)(30,15)%
 \Aqu(15,15)(7,0,180)
 \Aqu(15,15)(7,180,360)
}}
\def\selfMmaster{\pic{%
 \Lsc(0,15)(13,15)%
 \Lgl(17,15)(30,15)%
 \CCirc(15,15){5}{Black}{Gray}
}}
\def\selfMA{\pic{%
 \Lsc(0,15)(8,15)%
 \Lgl(22.5,15)(29.5,15)%
 \PhotonArc(15,15)(7,0,360){1.3}{11}
}}
\def\selfMB{\pic{%
 \Lsc(0,15)(8,15)%
 \Lgl(22,15)(30,15)%
 \Agh(15,15)(7,0,180)
 \Agh(15,15)(7,180,360)
}}
\def\selfMD{\pic{%
 \Lsc(0,15)(8,15)%
 \Lgl(22,15)(30,15)%
 \Asc(15,15)(7,0,180)
 \Asc(15,15)(7,180,360)
}}
\def\selfMF{\pic{%
 \Lsc(0,15)(8,15)%
 \Lgl(22,15)(30,15)%
 \Asc(15,15)(7,0,180)
 \Agl(15,15)(7,180,360)
}}
\def\selfMG{\pic{%
 \Lsc(0,15)(8,15)%
 \Lgl(22,15)(30,15)%
 \Aqu(15,15)(7,0,180)
 \Aqu(15,15)(7,180,360)
}}
\def\selfFmaster{\pic{%
 \Lqq(0,15)(13,15)%
 \Lqq(17,15)(30,15)%
 \CCirc(15,15){5}{Black}{Gray}
}}
\def\selfFA{\pic{%
 \Lqq(0,15)(8,15)%
 \Lqq(22,15)(30,15)%
 \Agl(15,15)(7,0,180)
 \Aqu(15,15)(7,180,360)
}}
\def\irredA{\picb{%
 \Asc(25,15)(15,5,175)%
 \Aqu(25,15)(15,185,270)%
 \Aqu(25,15)(15,270,355)%
 \Lgl(25,0)(25,30)
 \Line(0,14)(10,14)%
 \Line(0,16)(10,16)%
 \Line(40,14)(50,14)%
 \Line(40,16)(50,16)%
}}
\def\irredB{\picb{%
 \Agl(25,15)(15,0,90)%
 \Asc(25,15)(15,90,175)%
 \Aqu(25,15)(15,185,270)%
 \Aqu(25,15)(15,270,360)%
 \Lgl(25,0)(25,30)%
 \Line(0,14)(10,14)%
 \Line(0,16)(10,16)%
 \Line(40,15)(50,15)%
 \Line(48,13)(52,17)%
 \Line(48,17)(52,13)%
 \Line(50,14)(60,14)%
 \Line(50,16)(60,16)%
}}
\def\irredC{\picb{%
 \Asc(25,15)(15,5,90)%
 \Agl(25,15)(15,90,180)%
 \Aqu(25,15)(15,180,270)%
 \Aqu(25,15)(15,270,355)%
 \Lgl(25,0)(25,30)%
 \Line(-10,14)(0,14)%
 \Line(-10,16)(0,16)%
 \Line(-2,13)(2,17)%
 \Line(-2,17)(2,13)%
 \Line(0,15)(10,15)%
 \Line(40,14)(50,14)%
 \Line(40,16)(50,16)%
}}
\def\irredD{\picb{%
 \Agl(25,15)(15,0,180)%
 \Aqu(25,15)(15,180,270)%
 \Aqu(25,15)(15,270,360)%
 \Lgl(25,0)(25,30)%
 \Line(-10,14)(0,14)%
 \Line(-10,16)(0,16)%
 \Line(-2,13)(2,17)%
 \Line(-2,17)(2,13)%
 \Line(0,15)(10,15)%
 \Line(40,15)(50,15)%
 \Line(48,13)(52,17)%
 \Line(48,17)(52,13)%
 \Line(50,14)(60,14)%
 \Line(50,16)(60,16)%
}}
\def\irredE{\picb{%
 \Aaqu(25,15)(15,0,90)%
 \Agl(25,15)(15,90,180)%
 \Aqu(25,15)(15,180,270)%
 \Agl(25,15)(15,270,360)%
 \Lqu(25,0)(25,30)%
 \Line(-10,14)(0,14)%
 \Line(-10,16)(0,16)%
 \Line(-2,13)(2,17)%
 \Line(-2,17)(2,13)%
 \Line(0,15)(10,15)%
 \Line(40,15)(50,15)%
 \Line(48,13)(52,17)%
 \Line(48,17)(52,13)%
 \Line(50,14)(60,14)%
 \Line(50,16)(60,16)%
}}
\def\redA{\picb{%
 \Asc(25,15)(15,5,175)%
 \Aqu(25,15)(15,185,260)%
 \Aqu(25,15)(15,280,355)%
 \Line(0,14)(10,14)%
 \Line(0,16)(10,16)%
 \Line(40,14)(50,14)%
 \Line(40,16)(50,16)%
 \CCirc(25,0){5}{Black}{Gray}
}}
\def\redB{\picb{%
 \Asc(25,15)(15,5,175)%
 \Aqu(25,15)(15,185,355)%
 \Line(0,14)(10,14)%
 \Line(0,16)(10,16)%
 \Line(40,14)(50,14)%
 \Line(40,16)(50,16)%
 \CCirc(25,30){5}{Black}{Gray}
}}
\def\redC{\picb{%
 \Agl(25,15)(15,0,90)%
 \Asc(25,15)(15,90,175)%
 \Aqu(25,15)(15,185,360)%
 \Line(0,14)(10,14)%
 \Line(0,16)(10,16)%
 \Line(40,15)(50,15)%
 \Line(48,13)(52,17)%
 \Line(48,17)(52,13)%
 \Line(50,14)(60,14)%
 \Line(50,16)(60,16)%
 \CCirc(25,30){5}{Black}{Gray}
}}
\def\redD{\picb{%
 \Asc(25,15)(15,5,90)%
 \Agl(25,15)(15,90,180)%
 \Aqu(25,15)(15,180,355)%
 \Line(-10,14)(0,14)%
 \Line(-10,16)(0,16)%
 \Line(-2,13)(2,17)%
 \Line(-2,17)(2,13)%
 \Line(0,15)(10,15)%
 \Line(40,14)(50,14)%
 \Line(40,16)(50,16)%
 \CCirc(25,30){5}{Black}{Gray}
}}
\def\redE{\picb{%
 \Agl(25,15)(15,0,180)%
 \Aqu(25,15)(15,180,360)%
 \Line(-10,14)(0,14)%
 \Line(-10,16)(0,16)%
 \Line(-2,13)(2,17)%
 \Line(-2,17)(2,13)%
 \Line(0,15)(10,15)%
 \Line(40,15)(50,15)%
 \Line(48,13)(52,17)%
 \Line(48,17)(52,13)%
 \Line(50,14)(60,14)%
 \Line(50,16)(60,16)%
 \CCirc(25,30){5}{Black}{Gray}
}}
\def\redF{\picb{%
 \Agl(25,15)(15,0,180)%
 \Aqu(25,15)(15,180,260)%
 \Aqu(25,15)(15,280,360)%
 \Line(-10,14)(0,14)%
 \Line(-10,16)(0,16)%
 \Line(-2,13)(2,17)%
 \Line(-2,17)(2,13)%
 \Line(0,15)(10,15)%
 \Line(40,15)(50,15)%
 \Line(48,13)(52,17)%
 \Line(48,17)(52,13)%
 \Line(50,14)(60,14)%
 \Line(50,16)(60,16)%
 \CCirc(25,0){5}{Black}{Gray}
}}
\def\redX{\picb{%
 \Asc(25,15)(15,5,175)%
 \Aqu(25,15)(15,185,355)%
 \Line(0,14)(10,14)%
 \Line(0,16)(10,16)%
 \Line(40,14)(50,14)%
 \Line(40,16)(50,16)%
}}
\def\redY{\picb{%
 \Agl(25,15)(15,0,180)%
 \Aqu(25,15)(15,180,360)%
 \Line(-10,14)(0,14)%
 \Line(-10,16)(0,16)%
 \Line(-2,13)(2,17)%
 \Line(-2,17)(2,13)%
 \Line(0,15)(10,15)%
 \Line(40,15)(50,15)%
 \Line(48,13)(52,17)%
 \Line(48,17)(52,13)%
 \Line(50,14)(60,14)%
 \Line(50,16)(60,16)%
}}
\makeatletter \@addtoreset{equation}{section} \makeatother
\renewcommand{\theequation}{\arabic{section}.\arabic{equation}}
\makeatletter
\renewcommand\section{\@startsection {section}{1}{\z@}%
                                   {-5.5ex \@plus -1ex \@minus -.2ex}% bfr-
                                   {2.3ex \@plus.2ex}%
                                   {\normalfont\large\bfseries}}
\renewcommand\subsection{\@startsection{subsection}{2}{\z@}%
                                     {-3.25ex\@plus -1ex \@minus -.2ex}%
                                     {1.5ex \@plus .2ex}%
                                     {\normalfont\normalsize\bfseries}}
\renewcommand\thesection {\@arabic\c@section}
\renewcommand\thesubsection   {\thesection.\@arabic\c@subsection}
\renewcommand{\@seccntformat}[1]{%
\csname the#1\endcsname.\hspace{1.0em}}
\makeatother
\begin{document}

\flushbottom

\begin{titlepage}

\begin{flushright}
% OUTLINE  \\ 
% DRAFT \\ 
% arXiv:2203.05772\\ 
July 2022
\end{flushright}
\begin{centering}
\vfill

{\Large{\bf
 Sterile neutrino rates for general $M$, $T$, $\mu$, $k$: \\[3mm]
 review of a theoretical framework
}} 

\vspace{0.8cm}

M.~Laine % $^\rmi{a}$
 
\vspace{0.8cm}

% $^\rmi{a}$%
{\em
AEC, 
Institute for Theoretical Physics, 
University of Bern, \\ 
Sidlerstrasse 5, CH-3012 Bern, Switzerland \\}

\vspace*{0.8cm}

\mbox{\bf Abstract}
 
\end{centering}

\vspace*{0.3cm}
 
\noindent
The temperature of sphaleron freeze-out, $T\approx 130$~GeV, 
sets a scale for mechanisms for generating lepton 
asymmetries and converting them to baryon asymmetry (leptogenesis). 
For Majorana masses $M \gg \pi T$, leptogenesis takes place in the 
non-relativistic regime, while for $M \ll \pi T$, the dynamics 
is ultrarelativistic. The intermediate case $M \sim \pi T$ is 
the most cumbersome, as no effective theory methods are available. 
We review the definitions of and provide integral representations 
for all rate coefficients and mass corrections of $\mathcal{O}(h_\nu^2)$ 
in this regime, such that the expressions extrapolate to known limits 
and are gauge independent to the order computed, both in the symmetric
and the Higgs phase, for any helicity, momentum, and set 
of chemical potentials. Road signs towards a numerical 
evaluation are offered. 
%

%% %\noindent
%% %PACS numbers: 
%% %11.10.Wx, %        Finite temperature field theory
%% %11.15.Ha, %        Lattice gauge theory  
%% %12.38.Bx, %        Perturbative calculations in QCD
%% %12.38.Mh, %        Quark--gluon plasma
%% %14.40.Nd, %        Bottom mesons
%% %\\

\vspace*{0.8cm}

\noindent
\mbox{\bf Keywords:}
thermal field theory, neutrino physics, dark matter, leptogenesis

\vfill
 
\end{titlepage}

\tableofcontents

%%%%%%%%%%%%%%%%%%%%%%%%%%% SECTION %%%%%%%%%%%%%%%%%%%%%%%%%%%%%%%%%%%%%%
%
\section{Introduction}

The only concrete feature of known particle physics that is not
explained by the traditional version of the Glashow-Weinberg-Salam
Standard Model, is the appearance of neutrino masses, 
and the related phenomenon
that neutrino flavours mix with each other. Even 
if this shortcoming can be fixed in a simple way, by adding 
gauge singlet (sterile) right-handed neutrino fields in the theory, 
much remains unknown. In the active neutrino sector, 
major challenges are the determination of the absolute scale
of neutrino masses~\cite{katrin}
and of a CP-violating parameter $\delta$ that manifests itself
in their mixings~\cite{t2k}. 
In the sterile sector, 
the magnitude of the so-called Majorana 
mass parameters remains undetermined. 
In principle, they could vanish 
(in which case neutrinos are said to be Dirac-like), be as heavy as 
$10^{15}$~GeV, or be practically anything in between
(in the latter cases we talk about Majorana-like neutrinos). It would
be a major breakthrough
to establish the Majorana mass scale. 

In concrete terms, the Lagrangian of the Standard Model completed
by right-handed neutrino fields reads 
\be
 \mathcal{L}^{ }_\rmi{new-SM}  \equiv  
 \mathcal{L}^{ }_\rmi{old-SM} 
 + \bar{\nu}^{ }_\rmii{R} i \msl{\partial} \nu^{ }_\rmii{R} 
% \nn[0.5mm] 
% & - & 
 - 
 \bigl( 
 \bar{\nu}^{ }_\rmii{R}\, \tilde{\phi}_{ }^\dagger 
 h_{\nu}^{ }\, \ell^{ }_\rmii{L}
 + 
 \bar{\ell}^{ }_\rmii{L}\, h_{\nu}^\dagger \, 
 \tilde{\phi} \; \nu^{ }_\rmii{R} 
 \bigr)
% \nn[0.5mm] 
% & - & 
 - 
 \frac{1}{2}
 \bigl(
 \bar{\nu}_\rmii{R}^c M^{ }_\rmii{M} \nu^{ }_\rmii{R} 
 + 
 \bar{\nu}_\rmii{R}^{ }\, M^{\dagger}_\rmii{M} \nu^c_\rmii{R} 
 \bigr)
 \;, \la{L}
\ee 
where $\tilde{\phi}\, \equiv i \sigma^{ }_2 \phi^*$ is a conjugated
Higgs doublet and $\nu^c_\rmii{R}$ is a charge-conjugated right-handed
neutrino. 
The Majorana mass matrix $M^{ }_\rmii{M}$ is in general complex and
non-diagonal, but it necessarily satisfies 
$M^{ }_\rmii{M} = M^{T}_\rmii{M}$. 
A singular value decomposition 
permits to write it as 
$
 M^{ }_\rmii{M} = 
 O \mathop{\mbox{diag}}(M^{ }_1, M^{ }_2, M^{ }_3)\, O^T_{ }
$, 
where $M^{ }_\I \ge 0$ can be set in increasing order. In the following, 
we denoted a generic Majorana mass parameter by $M$. 

A possible indirect handle on physics in the neutrino sector can be offered
by cosmology. For instance, already several decades ago, it was 
realized that cosmological considerations require the existence of 
three light neutrino species. Today the corresponding
parameter, called $N^{ }_\rmi{eff}$, is precisely measured~\cite{planck}
and even more precisely computed 
theoretically~\cite{Neff6}. 
The agreement between these numbers is a great success of modern cosmology. 

There are also issues in cosmology that the traditional Standard Model
cannot explain. Two famous ones are the apparent existence of particle
dark matter, and of an asymmetry in the amounts of matter and antimatter
(the latter is referred to as baryon asymmetry). 
Astronomical observations have determined both parameters with 
percent level accuracy~\cite{planck}. 
If we are lucky, the theoretical explanations
of these features could be related to neutrino mass generation, and might 
allow for us to constrain the Majorana mass parameters. 

The possibility that the right-handed neutrino fields, and the corresponding
mass eigenstates, sterile neutrinos, could contribute to dark 
matter and baryon asymmetry, have been proposed long ago. Building on 
earlier works~\cite{bd,ekt}, 
it has been realized that for dark matter,
one of the sterile neutrinos should be sufficiently light ($M\sim $~keV)
and thus long-lived~\cite{dw,sf,review}. 
Even if unconfirmed, concrete demonstrations that 
the astronomical verification of 
this scenario could be possible, 
have been put forward~\cite{observe1,observe2}. 

For baryon asymmetry, processes originated
from sterile neutrinos are referred to as leptogenesis~\cite{classic}. 
Here, the
mass scale could be anything from $M\sim 0.1$~GeV~\cite{ars,as} to 
$M \sim 10^{15}$~GeV~\cite{classic}, 
with the lower bound again related to the 
well-constrained $N^{ }_\rmi{eff}$~(cf.,\ e.g.,\ 
refs.~\cite{Neff,Neff01,Neff05,Neff0} and references therein). 
Close to the lower bound, it is conceivable that sterile neutrinos
could be experimentally detectable in the future~\cite{ship}. 
In the full mass range, 
active neutrino masses are generated through
the see-saw mechanism~\cite{ss1,ss2,ss3}, which implies 
that the neutrino Yukawa couplings are small for small 
Majorana masses, spanning the range 
$|h^{ }_\nu| \sim 10^{-8\ldots 0}_{ }$. 
If $M \lsim 10^6$~GeV~\cite{davidsonibarra,turner}, 
some sterile neutrinos should be 
degenerate in mass, in order to permit for resonant enhancement
of CP violation that is necessary for generating the observed
baryon asymmetry~\cite{hambye,ap,tev}.  
The case $M\sim 100$~GeV, i.e.\ of the order of the weak scale, 
has attracted detailed attention only recently
(cf.,\ e.g.,\ refs.~\cite{interp0,interp1,interp2,interp3}
and references therein), even though it could arguably 
be the most ``natural'' one.

The goal of this paper is to assemble up-to-date
tools for addressing the cosmology of sterile
neutrinos in broad mass and temperature ranges. In particular we define and
derive rate coefficients and mass corrections that characterize
their interactions, oscillations and decoherence within 
a cosmological (or astrophysical) setting. The exposition is
somewhat theoretical in nature, as the numerical evaluation
poses challenges 
of its own~\cite{phasespace,lpm}; 
we return to an outlook at the end. 

The presentation is organized as follows. 
After recalling the general form of the rate equations satisfied
by right-handed neutrino density matrices and lepton asymmetries
(cf.\ \se\ref{se:eqs}), we review 
what has been known about the associated rate coefficients 
(cf.\ \se\ref{se:previous}). 
To proceed beyond this level, we 
discuss the Feynman rules pertinent to the problem
(cf.\ \se\ref{se:feynman}), given that the presence of chemical potentials 
leads to non-standard features. The $2\leftrightarrow 2$ and 
$1\leftrightarrow 3$ contributions to the rate coefficients
are analyzed in \se\ref{se:sterile_2to2}, 
whereas $1+n \leftrightarrow 2+n$ processes
are treated in \se\ref{se:sterile_1to2}. Thermal corrections
to dispersion relations, such as thermal masses, 
are the subject of \se\ref{se:mass}.
We conclude with a summary and outlook in \se\ref{se:x_concl}. 
Matrix elements squared relevant for 
the problem are listed in appendix~A. 

%%%%%%%%%%%%%%%%%%%%%%%%%%% SECTION %%%%%%%%%%%%%%%%%%%%%%%%%%%%%%%%%%%%%%
%
\section{Rate equations}
\la{se:eqs}

In order to study the cosmology of sterile neutrinos, and particularly
their contribution to dark matter and/or baryon asymmetry, a suitable
theoretical framework is needed. The early universe represents a 
multiparticle system, with a maximal temperature likely much higher
than any of the known particle masses. In this situation multiple
interactions take place. The traditional tool of 
particle physics, based on Feynman diagrams, describes individual
scatterings of a few particles, and is not trivially suited
to such a situation. Rather, 
it has to be combined with tools of statistical physics. 

An important notion in statistical physics is that of thermal equilibrium.
A system in thermal equilibrium is simple: it carries a minimal
amount of information (entropy is maximal), which can be characterized
by a handful of parameters, namely the temperature 
and the chemical potentials
associated with conserved charges. 

Of course, the universe as we observe it is {\em not} in thermal 
equilibrium. For instance, for dark matter, equilibrium would imply
that density is exponentially suppressed by 
$\sim (M T / \pi)^{3/2} e^{-M/T}$, 
which would lead to an energy density much smaller than 
the observed one, if $M\sim$~keV and $T\sim 10^{-4}$~eV. 
Baryon asymmetry cannot be generated in thermal equilibrium 
either, as famously stated by Sakharov~\cite{sakharov}. 

Even if the complete universe cannot be in thermal equilibrium, 
most of the Standard Model particles {\em were} in equilibrium
for most of the time~\cite{eR}. 
The key question for explaining dark matter and 
baryon asymmetry is to find particle types which decoupled from 
equilibrium at some point, or never entered it in the first place. 

A suitable framework for describing such dynamics is 
to classify degrees of freedom (that is, the densities
or density matrices of various particle 
species), as either ``fast'' or ``slow''. We call degrees of freedom
fast, if they experience reactions often enough to stay in thermal 
equilibrium. In contrast, slow variables can fall out of equilibrium. 

For fast variables, it is easy to determine their density
(or energy density), as it 
is completely characterized by the temperature and chemical potentials, 
with the distribution functions taking the Fermi-Dirac
or Bose-Einstein form. In contrast, for slow variables 
we do not know the density in advance. We need to carry out a 
theoretical computation to find it out. The computation is based
on a set of rate equations, which in turn are characterized by
the expansion rate of the universe (Hubble rate) as well as 
the most important quantities concerning us here, 
{\em thermal rate coefficients}. In general, this theoretical framework 
can be referred to as the Landau theory of thermodynamic 
fluctuations~\cite{landau5}. 

\vspace*{5mm}

Focussing on our problem, the most important slow variables are: 
\bi

\item
 lepton and baryon asymmetries carried by 
 Standard Model particles, 
 which in the following are denoted by 
 $n^{ }_a$ and $n^{ }_\B$, 
 with $a \in\{e,\mu,\tau\}$;

\item 
 two $3\times 3$ density matrices for the sterile neutrinos, 
 which in the following are denoted by 
 $
  \rho^{ }_\rmii{$(\pm)$}
 $,  
 or their linear combinations, defined 
 according to \eq\nr{symmetrization}. The subscripts 
 $(\pm)$ refer to two helicity states, as carried by
 massive spin-1/2 particles, and  
 the requirement for $3\times 3$ matrices originates from the 
 three different generations.

\ei
The evolutions of these variables are coupled to each other. 
We note in passing that at very high temperatures further
slow variables emerge~\cite{beneke}, but we restrict to the simplest
situation in the following, viable for $T \lsim 10^5$~GeV~\cite{eR}.  

The evolution equations for the slow variables 
are parametrized by a number of coefficients. 
One of the coefficients 
is a special one, namely the rate of anomalous baryon plus
lepton number violation that is present in 
the Standard Model~\cite{hooft,krs}
(conventionally this is referred to as the ``sphaleron rate''). 
In many leptogenesis scenarios, lepton number production continues
down to temperatures where the anomalous rate rapidly
switches off. Given that the anomalous rate plays an essential role
in converting lepton asymmetries to the physically observable 
baryon asymmetry, the precise features of the switch-off do 
need to be incorporated~\cite{freezeout}. Fortunately, this 
rate and its switch-off have been precisely determined with 
the help of numerical simulations~\cite{sphaleron}. This rate
coefficient is also ``universal'', in the sense that it is 
independent of the properties of the sterile neutrinos. 
In the following, we concentrate on the other rate coefficients. 

Unless the right-handed neutrinos are 
extremely heavy ($M \gg 10^{15}_{ }$~GeV), 
the see-saw formula suggests that the neutrino Yukawa couplings are small 
($| h^ {}_\nu | ^2 \ll 1$). The smallness of the couplings is the very 
reason why the sterile neutrinos are slow variables. 
In this situation, we can carry out a perturbative
expansion, with the first non-trivial order being 
$\rmO(h_\nu^2)$. There are important new effects at 
$\rmO(h_\nu^4)$, notably those related to CP-violation, however 
such effects often factorize into a product of 
contributions of $\rmO(h_\nu^2)$.

At $\rmO(h_\nu^2)$, the sterile neutrino 
rate coefficients are related to the 2-point
``retarded'' correlation function of the composite operator to 
which the sterile neutrinos couple. This correlation function 
is a matrix in Dirac space, and its matrix elements determine
the helicity components of the rate coefficients. The general
framework by which one arrives at this 2-point correlator is 
similar to a classic one considering active neutrino oscillations
in a statistical background~\cite{sr}, however a more general
derivation was needed, in order to properly include vacuum masses, 
both helicity states, as well as the fact that electroweak 
gauge bosons become light at high temperatures, 
and eventually the electroweak symmetry gets restored.  
Building on a number of previous steps, 
general derivations of the rate 
equations and definitions of the rate coefficients 
have been provided in refs.~\cite{simultaneous,dbX}. 

In order to specify the rate equations in a compact form, we consider
helicity-symmetrized or antisymmetrized density matrices,  
\be
 \rho^{\pm}_{ } 
 \; \equiv \; 
 \frac{\rho^{ }_\rmii{$(+)$} \pm \rho^{ }_\rmii{$(-)$}}{2}
 \;. \la{symmetrization} 
\ee
If the sterile neutrinos are not degenerate, the off-diagonal 
components of the density matrices average out,  
up to effects suppressed by $1/M$. 
 
Now, the evolution equations for lepton 
asymmetries take the form 
\be
 \dot{n}^{ }_a - \frac{\dot{n}^{ }_\B}{3}
  =  
%%%%%%%%%%%%% 
 4 \int_{\vec{k}} \tr \Bigl\{ 
 \bigl[ \rho^{+}_{ } - \nF^{ }(\omega^{ }_{ }) \bigr]
    {B}^{+}_{(a)} 
 + \rho^{-}_{ }\, {B}^{-}_{(a)}
 - \nF^{ }(\omega^{ }_{ }) \bigl[1 - \nF^{ }(\omega^{ }_{ })\bigr]\,
  {A}^+_{(a)}
 \Bigr\} \hspace*{4mm}
 \;,  \la{dna}
\ee
where 
$
 \int_{\vec{k}} \equiv \int 
 \! \frac{{\rm d}^3\vec{k}}{(2\pi)^3}
$, 
$\nF^{ }(\omega)$ denotes a matrix with Fermi distributions on the diagonal, 
and $\omega^{ }_{\I} \equiv \sqrt{k^2 + M_{\I}^2}$ are the 
corresponding frequencies.
The density matrices evolve as 
\ba
 \dot{\rho}^{\pm}_{ } & = & 
   i \bigl[
      {\omega}^{ }_{ }
    - {H}^{+}_{ }, \rho^{\pm}_{ }
   \bigr]
 \; - \;  
   i \bigl[
     {H}^{-}_{ }, \rho^{\mp}_{ }
   \bigr]
 \nn[2mm] 
 & + & 
 \bigl\{  
   {D}^{\pm}_{ } \,,\,
   \nF^{ }(\omega^{ }_{ }) - \rho^+_{ }
 \bigr\}
 \; - \;  
 \bigl\{  
   {D}^{\mp}_{ } \,,\,
   \rho^-_{ }
 \bigr\}
 \; + \; 
 2  \nF^{ }(\omega^{ }_{ })
 \bigl[ 1 - \nF^{ }(\omega^{ }_{ }) \bigr]
 \, {C}^{\pm}_{ }
 \;, \la{d_rhoH_1p2}  
\ea
where again $\omega$ is a diagonal matrix. 
The coefficients $A$ and $C$, appearing in the terms without density matrices, 
are proportional to chemical potentials, 
cf.\ \eqs\nr{matA} and \nr{matC}. 

After separating the neutrino Yukawa couplings into
C-even and C-odd parts, via 
\be
 \rephase \; \equiv \; \re ( h^{ }_{\I a}h^*_{\J a} )
 \;, \quad
 \imphase \; \equiv \; - i \im ( h^{ }_{\I a}h^*_{\J a} )
 \;, \la{yukawas}
\ee 
and doing the same in the dynamical parts $Q$ and $U$ of the rate
coefficients, defined via \eqs\nr{def_Q} and \nr{def_U}, 
the coefficients appearing in 
\eqs\nr{dna} and \nr{d_rhoH_1p2}
can be resolved as~\cite{simultaneous} 
%%%%%%%%%%%%%%%%%%%%%%%%%%%%%%%%%%%%%%%%%%%%%%%%%%%%%%%%%%%%%%%%%%%%%%%%%%%
\ba
 {A}^{+}_{(a)\I\I} & = &
   \bmu^{ }_a \phi^{+}_{(a)\I\I}
      {Q}^{+}_{(a)}
 \;, \la{matA} \\  
%%%%%
 {B}^{\pm}_{(a)\I\J} & = & 
     \phi^{\mp}_{(a)\I\J} {Q}^{\pm}_{(a)}
   + \phi^{\pm}_{(a)\I\J} \,{\!\bar{Q}}^{\pm}_{(a)}
 \;, \la{matB} \\
%%%%%%%%
 {C}^{\pm}_{\I\J} & = & 
      {\textstyle\sum_a}\, \bmu^{ }_a\, 
      \bigl[\,
      \phi^{\mp}_{(a)\I\J}
      {Q}^{\pm}_{(a)} 
    + \phi^{\pm}_{(a)\I\J}
     \,{\!\bar{Q}}^{\pm}_{(a)} 
     \,\bigr]
 \;, \la{matC} \\ 
%%%%%%%%
 {D}^{\pm}_{\I\J} & = & 
      {\textstyle\sum_a} 
    \bigl[\,
    \phi^{\pm}_{(a)\I\J} 
    {Q}^{\pm}_{(a)}
  + \phi^{\mp}_{(a)\I\J}\,
   {\!\bar{Q}}^{\pm}_{(a)} 
    \,\bigr]
 \;, \la{matD} \\
%%%%%%%%%%%%%%%%%
 {H}^{\pm}_{\I\J} & = & 
     {\textstyle\sum_a} 
   \bigl[\, 
   \phi^{\pm}_{(a)\I\J} 
   {U}^{\pm}_{(a)}
 + \phi^{\mp}_{(a)\I\J} 
   \,{\!\bar{U}}^{\pm}_{(a)} 
   \, \bigr]
 \;, \la{matH}  
\ea
where we have denoted leptonic chemical 
potentials by $\bmu^{ }_a \equiv \mu^{ }_a/T$.
The $Q$'s and $U$'s are absorptive (i.e.\ rate) 
and dispersive (i.e.\ mass) coefficients, respectively, 
capturing the dynamics of the heat bath. Specifically, like
in \eq\nr{symmetrization}, 
$
 {Q}^{\pm}_\rmii{$(a)$} = [{Q}^{ }_\rmii{$(a+)$}\pm
 {Q}^{ }_\rmii{$(a-)$}]/2
$ denote symmetrization and antisymmetrization with respect to helicity, 
and similarly for $U^{\pm}_\rmii{$(a)$}$. 
The values
$Q^{ }_\rmii{$(a\pm)$}$, 
$U^{ }_\rmii{$(a\pm)$}$,  
and the corresponding C-odd parts 
${\!\bar{Q}}^{ }_\rmii{$(a\pm)$}$, 
${\!\bar{U}}^{ }_\rmii{$(a\pm)$}$,  
are obtained from the imaginary and real parts of a retarded correlator, 
as defined in \eqs\nr{def_Q} and \nr{def_U}. 

%%%%%%%%%%%%%%%%%%%%%%%%%%% SECTION %%%%%%%%%%%%%%%%%%%%%%%%%%%%%%%%%%%%%%
%
\section{What is currently known and not known about rate coefficients}
\la{se:previous}

The dynamical parts of the rate coefficients, denoted by $Q$ and $U$ above, 
originate from matrix elements of a 2-point correlator. 
Their definitions will be given  
in \eqs\nr{def_Q} and \nr{def_U}, 
once the notation needed has been introduced. 
Given the definitions, the correlator and its matrix elements 
need to be evaluated.
This task has only been completed 
in a number of special cases. 
The status strongly depends on the ``regime'' considered, 
i.e.\ the relation of the 
mass of a sterile neutrino ($M$) and the temperature ($T$). 
In thermal field theory, the temperature naturally appears 
in the combination $\pi T$.
The following regimes can be identified: 
\bi

\item
{\em non-relativistic regime}, 
$M \gg \pi T$.
In this situation thermal corrections are small. Actually, there has been
a long discussion about how small they are. 
Based on leading-order computations, 
one might naively think that thermal corrections
are Boltzmann-suppressed, i.e.\ exponentially small. It turns out, however,
that in general thermal corrections are only power-suppressed~\cite{ope}. 

\item
{\em relativistic regime}, 
$M\sim \pi T$.
In this situation thermal corrections are of $\rmO(1)$. 
Then they must be determined without approximation, 
leading to cumbersome computations. 

\item
{\em ultrarelativistic regime}, 
$M \ll \pi T$.
In this situation thermal corrections dominate the dynamics, and physics
looks very different from that in vacuum. Fortunately, a lot of work has
been carried out for this situation in the context of QCD, and 
effective field theory methods and resummations have been developed, 
notably an ingenious simplification relevant 
for light-cone correlators, originating from considerations
of causality in a thermal setting~\cite{sch}. 

\ei

Beyond order-of-magnitude estimates or 
leading-order computations in the non-relativistic regime 
(cf., e.g., refs.~\cite{riotto,bumuwi,sacha,sb_rev,bb_rev} for reviews), 
the state of the art of the determination of the rate coefficients 
in the different domains can be summarized as follows: 
\bi

\item
in the {\em non-relativistic regime}, the simplest CP-even
rate coefficient, which is of $\rmO(h_\nu^2)$ in neutrino 
Yukawa couplings, has been determined up to next-to-leading 
order (NLO) in Standard Model 
couplings~\cite{salvio,nonrel,biondini,kubo,sangel}, 
partly by making use of the methodology of ref.~\cite{ope}. 
Subsequently, the simplest CP-odd rate coefficient, 
which is of $\rmO(h_\nu^4)$ in neutrino Yukawa couplings,  
has also been determined up to NLO in Standard Model 
couplings in the non-relativistic
regime~\cite{db1,nora,db2,racker}.

\item
in the {\em relativistic regime}, 
the NLO level has been reached  
for the simplest CP-even rate coefficient
in the symmetric phase~\cite{bg,relativistic}.

\item
in the {\em ultrarelativistic regime}, 
the simplest CP-even rate coefficient has been determined
to full leading order (LO) 
in the symmetric phase~\cite{bb1,bb2}, which in this case 
requires a systematic resummation of the loop expansion.

\item
a smooth {\em interpolation} between NLO results in the
non-relativistic and relativistic regimes and resummed LO results
in the ultrarelativistic regime has been been worked out for
the simplest CP-even rate coefficient
in the symmetric phase~\cite{interpolation}. 

\item
the above results have been generalized in a number of ways, 
to include chemical potentials~\cite{n3}, 
helicities~\cite{lello,cptheory}, 
as well as the Higgs (or ``broken'') phase~\cite{broken,degenerate}.

\item
making use of the methods of ref.~\cite{sch}, the NLO level
has been reached for the first rate coefficient even in the
ultrarelativistic regime
in the Higgs phase~\cite{nlo_width}.

\ei

Despite this progress,  
open goals remain. A few shortcomings 
that need to be overcome to permit for phenomenological
studies in the full parameter space, are as follows:  
\bi

\item[(i)] 
 in the relativistic regime,
 only the sum over helicity states has been addressed at NLO; 
 the difference between 
 {\em helicity-conserving and helicity-flipping rates} has
 only been determined in the ultrarelativistic regime. 
 This is a problem, given that helicity asymmetry
 has the same quantum numbers as the lepton asymmetry, 
 and therefore plays an important role in the rate equations. 
 (Extrapolations towards the relativistic regime, however without
 a full computation, have been presented in ref.~\cite{interp1}.)
 
\item[(ii)]
 in the relativistic regime, 
 the {\em chemical potential dependence} of the rate 
 coefficients has not been resolved at NLO.
 This needs to be done, given that
 chemical potentials are proportional to lepton asymmetries, and therefore
 give again a formally order unity contribution to the rate equations.

\item[(iii)]
 in the relativistic regime, 
 rate coefficients have been
 systematically determined only in the symmetric phase at NLO. 
 Their determination becomes much more complicated in the {\em Higgs phase}, 
 where Standard Model masses play a role. This is a shortcoming
 for leptogenesis computations, given that the electroweak crossover 
 is at $T \approx 160$~GeV~\cite{crossover,dono} whereas the sphaleron
 rate switches off at $T \approx 130$~GeV~\cite{sphaleron}, 
 i.e.\ on the side of the Higgs phase.
 Moreover, lepton asymmetry generation 
 can continue at $T < 130$~GeV~\cite{singlet,late,eijima,simultaneous,new},  
 and this may play an important role 
 for sterile neutrino dark matter, 
 given that large lepton asymmetries need to be present at 
 $T \sim 0.2$~GeV when dark matter production 
 peaks~\cite{sf,shifuller,dmpheno,coincident,dbY}. 

\item[(iv)]
 apart from a few test cases~\cite{kinetic,cpnumerics,degenerate}, 
 rate coefficients have been evaluated after momentum averaging
 (cf.,\ e.g.,\ refs.~\cite{old1,old2,asaka,ht,new1,new2,new3,new4} 
 for recent works). 
 This assumes that density matrices
 depend on the momentum $k$ through the shape of the Fermi distribution, 
 with only their overall amplitude appearing
 as a dynamical variable. Even if 
 often accurate up to a factor of order unity, 
 discrepancies of an order of magnitude are also possible~\cite{degenerate}, 
 so the {\em full momentum dependence} of 
 the rate coefficients needs to be established. In the relativistic
 regime, this has so far only been achieved for the simplest CP-even
 rate coefficient in the symmetric phase at NLO. 

\item[(v)]
 the {\em gauge independence} of the NLO rate coefficients has not been 
 demonstrated in the Higgs phase, and the possibility to separate their 
 origin into ``direct'' and ``indirect'' contributions has been 
 asserted~\cite{broken} but not properly proven.  

\ei

\vspace*{5mm}

At this point, it is helpful to be more specific about the order that
needs to be reached for phenomenologically relevant computations. Above, 
NLO results were mentioned for the non-relativistic and relativistic 
regimes. In these regimes, the LO reactions are $1\leftrightarrow 2$
processes, such as the decays of the sterile neutrinos into 
Standard Model particles or, if the sterile neutrinos are light, 
the decays of Higgs or gauge bosons into sterile neutrinos and 
Standard Model leptons. NLO reactions then comprise of 
virtual corrections to $1\leftrightarrow 2$ processes, supplemented
by new classes of real processes, notably 
$2\leftrightarrow 2$ and $1\leftrightarrow 3$ scatterings. 

Now, when we go to the ultrarelativistic regime
($m^{ }_i,M \ll \pi T$), the picture changes. 
The $1\leftrightarrow 2$ processes become phase-space suppressed
compared with the scaling dimension~$T$, whereas
$2\leftrightarrow 2$ reactions experience no suppression. The phase-space
suppression of $1\leftrightarrow 2$ processes also implies that they 
are sensitive to small corrections, necessitating 
the summation of soft $1+n\leftrightarrow 2+n$ scatterings, 
with $n\ge 0$. 
The upshot is that a full LO computation in the ultrarelativistic
regime must include all $2\leftrightarrow 2$ reactions~\cite{bb2} as well as
$1+n\leftrightarrow 2+n$ scatterings in a resummed form~\cite{bb1}. 
The latter goes under the 
name of Landau-Pomeranchuk-Migdal (LPM) 
resummation~\cite{gelis3,amy1,agmz}.

What is of interest to us is to obtain a result which is LO accurate
in all domains, and interpolates smoothly between them~\cite{interpolation}. 
It then needs to include $1\leftrightarrow 3$, 
$2\leftrightarrow 2$ and resummed
$1+n\leftrightarrow 2+n$ scatterings, in a way that that thermal mass 
corrections are included where necessary, 
but double countings are avoided. 
General numerical methods for determining the
$1\leftrightarrow 3$ and $2\leftrightarrow 2$ and 
the virtual corrections to $1\leftrightarrow 2$ processes 
that cancel mass singularities, have been developed 
in ref.~\cite{phasespace}, whereas a systematic approach
to $1+n\leftrightarrow 2+n$ scatterings and the interpolation
between the ultrarelativistic and relativistic regimes, 
has been put forward in ref.~\cite{lpm}.
In the remainder of this paper, we turn to how 
the ingredients necessary for refs.~\cite{phasespace,lpm} 
can be determined, for general $M,T,\mu,k$.

%%%%%%%%%%%%%%%%%%%%%%%%%%% SECTION %%%%%%%%%%%%%%%%%%%%%%%%%%%%%%%%%%%%%%
%
\section{Feynman rules for Standard Model in 
a statistical background}
\la{se:feynman}

In order to compute all rates, we need to 
establish the Feynman rules relevant to the problem. 
The way to include chemical potentials cannot easily be located in
literature so, for completeness, the ingredients are
outlined in \se\ref{ss:Rxi}, following ref.~\cite{degenerate}
but generalizing to an arbitrary gauge parameter. 
Subsequently, definitions of the dynamical coefficients~$Q$ and~$U$, 
alluded to in \se\ref{se:eqs}, are motivated in \se\ref{ss:observable}. 

%%%%%%%%%%%%%%%%%%%%%%%%%%% SUBSECTION %%%%%%%%%%%%%%%%%%%%%%%%%%%%%%%%%%%
%
\subsection{$R^{ }_\xi$ gauge in the presence of chemical potentials}
\la{ss:Rxi}

The presence of chemical potentials induces non-zero expectation
values for temporal gauge field components, 
so we need to include these
backgrounds when deriving Feynman rules. Within the imaginary-time formalism, 
we denote the gauge backgrounds by
\be
 \muY^{ } \;\equiv\; -i g^{ }_1 B^{ }_0 \;, \quad
 \muA^{ } \;\equiv\; -i g^{ }_2 A^3_0
 \;, \la{muYA}
\ee
where $B^{ }_\mu$ is a hypercharge field 
(cf.\ \eq\nr{Dmu} for sign conventions) 
and $g^{ }_1, g^{ }_2$ are the 
U$^{ }_\rmii{Y}(1)$ and SU$^{ }_\rmii{L}(2)$
gauge couplings, respectively. 

If we find ourselves on the side of the Higgs phase, 
it is useful to ``diagonalize'' the 
chemical potentials used, via a linear transformation.  
To this end we may write 
$
 \muY^{ } = \muQ^{ } + s^2 \muZ^{ }
$, 
$
 \muA^{ } = -\muQ^{ }+ (1-s^2) \muZ^{ }
$, 
where $s \equiv \sin(\tilde\theta)$, 
with $\tilde\theta$ denoting the weak mixing
angle associated with temporal gauge field 
components~\cite{degenerate}.\footnote{%
 The notation $\tilde\theta$ is introduced in order to make a distinction
 to the standard mixing angle $\theta$ that plays a role for the {\em spatial}
 gauge field components. 
 } 
After the transformation, 
$\muZ^{ }$ couples to the neutral weak current, 
whereas $\muQ^{ }$ is the chemical potential 
associated with electromagnetism.

The values of $\muY^{ }$ and $\muA^{ }$ are to be found dynamically, 
by extremizing an effective potential 
$V(v,\muY^{ },\muA^{ })$~\cite{khlebnikov}.
The solution implies that when the symmetry gets restored ($v \ll T$),
$\muA^{ }\to 0$, $s\to 0$, and $\muZ^{ }\to\muQ^{ }$.  
In contrast, deep in the Higgs phase ($v \gg T$), 
$\muZ^{ } \sim \muQ^{ }\,T^2/v^2 \ll \muQ^{ }$ 
and $\muY^{ } + \muA^{ } \approx 0$. 
As discussed in \se{3} of ref.~\cite{degenerate}, it is practical 
in the Higgs phase to choose $\muZ^{ } = 0$ in the tree-level 
Feynman rules, and let its small non-zero value be generated by 
1-loop tadpole corrections, at the same stage when 1-loop corrections
to dispersion relations are derived
(cf.\ \eq\nr{muZ}).
The tree-level chemical potentials induced 
by the gauge backgrounds to different Standard Model 
particles are listed in table~\ref{table:mus}. 

%%%%%%%%%%%%%%%%%%%%% TABLE %%%%%%%%%%%%%%%%%%%%%%%%%%%%%%%%%%%%%
%
\begin{table}[t]

{ %\small{
\begin{center}
\begin{tabular}{lclcl}
 \hline\hline \\[-5mm]
    & & left-handed state & & right-handed state \\
 \hline \hline \\[-3mm]
 up-type quarks & &
 $\displaystyle \mu_{u_{\rmii{$L$}}} \equiv \frac{\muB}{3} + \frac{\muY}{6}
 - \frac{\muA}{2}$
 & &
 $\displaystyle \mu_{u_{\rmii{$R$}}} \equiv \frac{\muB}{3} + \frac{2\muY}{3} $ 
 \\[3mm]
%%%%%%%%%%%%%%
 down-type quarks & & 
 $\displaystyle \mu_{d_{\rmii{$L$}}} \equiv \frac{\muB}{3} + \frac{\muY}{6}
 + \frac{\muA}{2}$
 & &
 $\displaystyle \mu_{d_{\rmii{$R$}}} \equiv \frac{\muB}{3} - \frac{\muY}{3} $ 
 \\[3mm]
%%%%%%%%%%%%%%
 neutrinos of flavour $a$ & & 
 $\displaystyle \munLa \equiv \mu^{ }_a - \frac{\muY}{2}
 - \frac{\muA}{2}$
 & &
 \\[3mm]
%%%%%%%%%%%%%%
 charged leptons of flavour $a$ & & 
 $\displaystyle \mueLa \equiv \mu^{ }_a - \frac{\muY}{2}
 + \frac{\muA}{2}$
 & &
 $\displaystyle \mu_{e_{\rmii{$R$}a}} \equiv \mu^{ }_a - \muY^{ } $ 
 \\[2mm]
  \hline \hline \\[-5mm]
  & & scalar & & gauge field \\
 \hline \hline \\[-3mm]
%%%%%%%%%%%%%%
 neutral  & & 
 $\displaystyle \muphin \equiv \frac{\muY}{2}
 + \frac{\muA}{2}$
 & &
 $\displaystyle \muZn \equiv 0$
 \\[3mm]
%%%%%%%%%%%%%%
 charged  & & 
 $\displaystyle \muphip \equiv \frac{\muY}{2}
 - \frac{\muA}{2}$
 & &
 $\displaystyle \muW^{ } \equiv -\muA^{ }$
 \\[3mm]
  \hline \hline 
%%%%%%%%%%%%%%%%%%%%%%%%%%%%% 
\end{tabular} 
\end{center}
}

\vspace*{3mm}

\caption[a]{\small
  Effective chemical potentials carried by 
  Standard Model particles at tree level in the chiral limit. 
  Here $\muB$ is the baryon chemical potential, 
  $\muY$ and $\muA$ are defined in \eq\nr{muYA},
  and $\mu^{ }_a$ is the chemical potential
  associated with the active lepton flavour $a\in\{e,\mu,\tau\}$. 
 }
\label{table:mus}
\end{table}
%
%%%%%%%%%%%%%%%%%%%%%%%%%%%%%%%%%%%%%%%%%%%%%%%%%%%%%%%%%%%%%%%%%%%%%

As a powerful crosscheck of the computations, 
it is useful to 
carry them out in a general~$R^{ }_{\xi}$ gauge, 
which needs to include gauge field backgrounds. 
With the conventions
\be
 D^{ }_\mu\, \phi = 
 \biggl( \partial^{ }_\mu + \frac{i g^{ }_1 B^{ }_\mu}{2}
                          - \frac{i g^{ }_{2\,} \sigma^a A^a_\mu}{2}
 \biggr)\phi
 \;, \quad
 \phi = 
 \biggl(
        \begin{array}{c} \phi^{ }_+ \\ \phi^{ }_0 \end{array} 
 \biggr)
 = 
 \frac{1}{\sqrt{2}}
 \biggl(
        \begin{array}{c} \phi^{ }_2 + i \phi^{ }_1 \\
                                  h - i \phi^{ }_3 \end{array} 
 \biggr)
 \;, \la{Dmu} 
\ee
where $\sigma^a_{ }$ are the Pauli matrices, the gauge fixing term 
appearing in the imaginary-time action reads
\be
 S \supset \int_X \frac{1}{2\xi} \sum_{a=0}^3 G_a^2 
 \;, 
\ee
where ($\mB^{ }\equiv g^{ }_1 v/2$, $\mW^{ }\equiv g^{ }_2v/2$)
\ba
 G^{ }_0 & \equiv & 
 \partial^{ }_\mu B^{ }_\mu - \xi \mB \phi^{ }_3
 \;, \\ 
 G^{ }_1 & \equiv & 
 \partial^{ }_\mu A^1_\mu - i \muA^{ } A^2_0 - \xi \mW \phi^{ }_1
 \;, \\ 
 G^{ }_2 & \equiv & 
 \partial^{ }_\mu A^2_\mu + i \muA^{ } A^1_0 - \xi \mW \phi^{ }_2
 \;, \\ 
 G^{ }_3 & \equiv & 
 \partial^{ }_\mu A^3_\mu  - \xi \mW \phi^{ }_3
 \;. 
\ea

When Feynman rules and ghost properties are derived from 
these gauge conditions, the usual Feynman rules are recovered, 
except that charged Goldstone modes and ghost fields carry 
the same chemical potentials as the $W^{\pm}_{ }$ gauge bosons, 
and the momenta appearing in vertices are shifted by chemical potentials. 
Defining 
\be
 \tilde P^{ }_{a_x} \;\equiv\; (p^{ }_n + i\mu^{ }_{a_x},\vec{p})
 \;, \la{tildeP}
\ee
where $p^{ }_n$ is a Matsubara frequency and $\mu^{ }_{a_x}$ 
a chemical potential associated with 
a particle of type $a^{ }_x$, the upshot
is that Feynman rules are like without chemical potentials, but 
with all momenta appearing in the shifted form 
$\tilde P^{ }_{a_x}$. Chemical equilibrium 
requires that $\sum_{x} \tilde P^{ }_{a_x} = 0$ at each vertex, 
so momentum conservation can be used as before. It should be noted, 
however, that the substitution $P\to -P$ implies that the sign
of a chemical potential gets inverted.

%%%%%%%%%%%%%%%%%%%%%%%%%%% SUBSECTION %%%%%%%%%%%%%%%%%%%%%%%%%%%%%%%%%%%
%
\subsection{Sterile neutrino correlator and its helicity projections}
\la{ss:observable}

At $\rmO(h_\nu^2)$, 
the coefficients parametrizing the rate equations of \se\ref{se:eqs}
are all related to the real and imaginary parts of a particular 
retarded correlation function,
\be
 \Pi^\rmii{R}_{a}(\K)
 \; \equiv \;
 \int_{\mathcal{X}} e^{i\K\cdot\mathcal{X}}
 \bigl\langle
   i \theta(t)
   \bigl\{
     (\tilde\phi^\dagger \ell^{ }_{\sL a})(\mathcal{X}) 
     \,,\, 
   (\bar{\ell}^{ }_{\sL a} \tilde\phi )(0)
   \bigr\}
  \bigr\rangle
 \la{PiR}
 \;.
\ee
We represent the retarded 
correlator as an analytic continuation of an imaginary-time one, 
\ba
 \Pi^\rmii{R}_{a}(\K)
 & = & 
 \Pi^\rmii{E}_{a}(\tilde{K}) 
 \bigr|^{ }
       _{\tilde{k}^{ }_n \to -i [\omega + i0^+]}
 \;, 
 \la{PiE} 
 \\[2mm]
%%%%%%%%%%%%%%%%%%%%
 \Pi^\rmii{E}_{a}(\tilde{K}) 
 & \equiv &  
 \int_X \! e^{i\tilde{K}\cdot X}
 \, 
 \bigl\langle \,
 (\tilde\phi^\dagger \ell^{ }_{\sL a})(X) \, 
 (\bar{\ell}^{ }_{\sL a} \tilde\phi )(0) \,
 \bigr\rangle 
 \;. \la{observable} 
\ea
Here $a\in\{e,\mu,\tau\}$ labels 
an active lepton flavour, 
$\mu^{ }_a$ is the corresponding chemical potential, 
$
 \K \cdot \mathcal{X}
 \equiv
 \omega t - \vec{k}\cdot\vec{x}
$, 
$
 \omega = \sqrt{k^2 + M^2}
$,
$
 \tilde K\cdot X = \tilde k^{ }_n \tau - \vec{k}\cdot\vec{x}
$, 
and 
$
 \tilde{k}^{ }_n = k^{ }_n -i \mu^{ }_a
$. 

For future reference, let us rewrite 
\eq\nr{observable} more explicitly, after 
going to the Higgs phase ($v\simeq 246$~GeV) 
and to momentum space. In a compact notation, 
implying a sum-integral over the four-momenta $P,Q$
in products where they appear, this yields
\ba
 \Pi^\rmii{E}_{a}(\tilde{K}) & = & 
 \frac{1}{\Omega}\, \biggl\langle 
  \frac{v^2}{2} \, \nu^{ }_{\sL a}(-\tilde{K}) 
                \, \bar{\nu}^{ }_{\sL a}(-\tilde{K})
 \nn[2mm] 
 & + &  \frac{v}{\sqrt{2}}
   \Bigl\{
%    \bigl[
            \phi^{ }_0(-\tilde{K} - \tilde{P})
            \, \nu^{ }_{\sL a}(\tilde{P})
            \, \bar{\nu}^{ }_{\sL a}(-\tilde{K})
          -
           \phi^{ }_+(-\tilde{K} - \tilde{P})
           \, e^{ }_{\sL a}(\tilde{P})
           \, \bar{\nu}^{ }_{\sL a}(-\tilde{K})
%    \bigr] 
%   \Bigr\}
 \nn 
 &  & 
        \hspace*{6mm} + \, 
%    \bigl[
           {\nu}^{ }_{\sL a}(-\tilde{K}) \,
           \bar{\nu}^{ }_{\sL a}(\tilde{P}) \, 
           \phi^{*}_0(-\tilde{K} - \tilde{P}) 
          -
           {\nu}^{ }_{\sL a}(-\tilde{K}) \,
           \bar{e}^{ }_{\sL a}(\tilde{P}) \, 
           \phi^{*}_+(-\tilde{K} - \tilde{P})
%    \bigr] 
   \Bigr\}
 \nn[2mm] 
 & + &
    \Bigl\{
            \phi^{ }_0(-\tilde{K} - \tilde{P})
            \, \nu^{ }_{\sL a}(\tilde{P})
          -
            \phi^{ }_+(-\tilde{K} - \tilde{P})
            \, e^{ }_{\sL a}(\tilde{P})
    \Bigr\}
 \nn & \times &   
    \Bigl\{ 
           \bar{\nu}^{ }_{\sL a}(\tilde{Q}) \, 
           \phi^{*}_0(-\tilde{K} - \tilde{Q}) 
          -
           \bar{e}^{ }_{\sL a}(\tilde{Q}) \, 
           \phi^{*}_+(-\tilde{K} - \tilde{Q})
    \Bigr\}
 \biggr\rangle 
 \;, \la{observable2}
\ea
where $\Omega = \int_X$ is the space-time volume. The first structure, 
proportional to $v^2$, represents an ``indirect'' contribution in 
the language of ref.~\cite{broken}. The last structure, containing
no~$v$, is a ``direct'' contribution, as the operator couples directly
to propagating modes. The middle structure in \eq\nr{observable2} 
represents a cross term between these two possibilities; 
the corresponding diagrams can be found in \fig\ref{fig:retarded}. 

Helicity matrix elements are obtained from 
the correlator $ \Pi^\rmii{R}_{a} $ as 
\be
 \bar{u}^{ }_{\vec{k}\tau} \, \aL  \Pi^\rmii{R}_{a} \, \aR 
   \, u^{ }_{\vec{k}\tau}
 = 
 \frac{ 
  \tr \bigl\{ 
         \bigl( \bsl{\K} + M \bigr)
         \eta^{ }_\tau \bar{\eta}^{ }_\tau 
         \bigl( \bsl{\K} + M \bigr)
         \aL \Pi^\rmii{R}_{a} \, \aR 
      \bigr\}
 }{\omega + M} 
 \;, \la{M_element}
\ee
where $\aL,\aR$ are chiral projectors, 
and we inserted on-shell spinors in the form
\be 
 u^{ }_{\vec{k}\tau} = 
 \frac{(\bsl{\K} + M)\,\eta^{ }_\tau}{\sqrt{\omega + M}} 
 \;, \quad \tau = \pm
 \;. 
\ee
Coefficients denoted by $Q$ and $\bar{Q}$ in \se\ref{se:eqs}
parametrize C-even and C-odd parts, respectively, 
of the ``absorptive'' part of the matrix element, normalized
by the energy, 
\be
 \frac{
  \im \bar{u}^{ }_{\vec{k}\tau}
         \aL \Pi^\rmii{R}_{a} \, \aR 
     u^{ }_{\vec{k}\tau} 
   }{\omega}
 \; \equiv \;  \left.  Q^{ }_{(a\tau)} \right|^{ }_\rmi{C-even}
 + \left. \bar{Q}^{ }_{(a\tau)} \right|^{ }_\rmi{C-odd} 
 \;, \la{def_Q}
\ee
whereas $U$ and $\bar{U}$ parametrize its 
``dispersive'' part,
\be
 \frac{
  \re \bar{u}^{ }_{\vec{k}\tau}
         \aL \Pi^\rmii{R}_{a} \, \aR 
     u^{ }_{\vec{k}\tau} 
   }{\omega}
 \; \equiv \;  \left.  U^{ }_{(a\tau)} \right|^{ }_\rmi{C-even}
 + \left. \bar{U}^{ }_{(a\tau)} \right|^{ }_\rmi{C-odd} 
 \;. \la{def_U}
\ee 
The C-even and C-odd parts can be extracted by symmetrizing and 
antisymmetrizing in chemical potentials, respectively.
(Note that neutrino Yukawa couplings have been factored out from
the Standard Model correlator.)

In order to simplify \eq\nr{M_element}, it is helpful to consider 
the sum and the difference of the helicity states. 
Resolving $\eta^{ }_\tau$ in terms of 2-component spinors, defined
as eigenstates of $\vec{k}\cdot\vec{\sigma}$, 
the sum is seen to amount to 
\be
 \sum_{\tau = \pm} \eta^{ }_{\tau}\bar{\eta}^{ }_\tau 
 = \frac{ \mathbbm{1} + \gamma^0_{ }}{2}
 \;. \la{pre_heli_sum}
\ee
Making use of the Clifford algebra we then get 
\be
 \sum_{\tau = \pm }
 \bar{u}^{ }_{\vec{k}\tau}
 \, \aL  \Pi^\rmii{R}_{a} \, \aR \, u^{ }_{\vec{k}\tau}
 = \tr\bigl\{ \bsl{\K} \aL \Pi^\rmii{R}_{a} \, \aR \bigr\}
 \;, \la{heli_sum}
\ee
where a mass term was eliminated by the chiral projectors. 

It is a bit less standard to work out the helicity difference. Making
use of an explicit (Dirac or Weyl) representation of the Dirac matrices, 
it can be verified that 
\be
 \eta^{ }_+ \bar{\eta}^{ }_+ - \eta^{ }_-\bar{\eta}^{ }_-
 = 
 \frac{(\mathbbm{1} + \gamma^{0}_{ })\,\gamma^{ }_5 \bsl{\K}
       (\mathbbm{1} + \gamma^{0}_{ })}{4 k}
 \;. \la{pre_heli_diff}
\ee
Inserting into \eq\nr{M_element}, and dropping again terms eliminated
by chiral projectors, yields
\be
 \sum_{\tau = \pm } \tau\,
 \bar{u}^{ }_{\vec{k}\tau} 
 \, \aL  \Pi^\rmii{R}_{a} \, \aR \, u^{ }_{\vec{k}\tau}
 = 
 \frac{ \tr\bigl\{ \bigl( \omega \bsl{\K} - M^2 \gamma^{0}_{ }\bigr)
                  \, \aL \Pi^\rmii{R}_{a} \, \aR \bigr\} }{k}
 \;. \la{heli_diff}
\ee

As a final step, the sum in \eq\nr{heli_sum} and the difference
in \eq\nr{heli_diff} can be employed for determining the positive and
negative helicity contributions. 
To summarize, if we define 
\be
 \Theta^{ }_{\E} \; \equiv \; 
 \tr\bigl\{ \msl{\E} \aL \Pi^\rmii{R}_{a} \, \aR \bigr\}
 \;, \la{Theta_E}
\ee
and work in the medium rest 
frame, denoting its four-velocity by $\U \equiv (1,\vec{0})$, 
then the helicity matrix elements are given by
\be
 \bar{u}^{ }_{\vec{k}\tau} 
 \, \aL  \Pi^\rmii{R}_{a} \, \aR \, u^{ }_{\vec{k}\tau}
 \; = \;
         \frac{(k + \tau\omega) \Theta^{ }_{\K}
      - \tau M^2 \Theta^{ }_{\U}} {2 k }
 \; = \; 
  \frac{
  \bigl( 
    \omega + \tau k
  \bigr)
    \Theta^{ }_{(1,\tau {\vec{k}} / {k} )}
  }{2}
% \;, \quad
% \vec{e} \equiv \Bigl( 0,  \Bigr)
% \;, \quad
% \tau = \pm
 \;. \la{hel_M}
\ee

As an example, at leading order in the Standard Model, 
$
 \Pi^\rmii{R}_{a} \supset 2 \aL \bsl{\P} \aR \, \mathcal{C}
$
(cf.\ \se\ref{ss:splitup_1to2}). 
It then follows that  
$
 \Theta^{ }_{\E} \supset 4 \E \cdot \P \, \mathcal{C} 
$, 
and correspondingly the positive helicity component 
experiences the rate  
$
 \bar{u}^{ }_{\vec{k} + } 
 \, \aL  \Pi^\rmii{R}_{a} \, \aR \, u^{ }_{\vec{k} +}
 = 
 2 (\omega + k ) (\epsilon - p^{ }_{\parallel})\, \mathcal{C}
$, 
where we wrote $\P \equiv (\epsilon,\vec{p})$ and 
$p^{ }_{\parallel} \equiv \vec{p}\cdot\vec{k} / k$
for the component parallel to $\vec{k}$.
Similarly, the negative helicity component yields
$
 \bar{u}^{ }_{\vec{k} - } 
 \, \aL  \Pi^\rmii{R}_{a} \, \aR \, u^{ }_{\vec{k} -}
 = 
 2 (\omega - k ) (\epsilon + p^{ }_{\parallel})\, \mathcal{C}
$.

%%%%%%%%%%%%%%%%%%%%%%%%%%% SECTION %%%%%%%%%%%%%%%%%%%%%%%%%%%%%%%%%%%%%%
%
\section{Rates from 
${2}\leftrightarrow{2}$ and ${1}\leftrightarrow{3}$ scatterings}
\la{se:sterile_2to2}

Moving on to the evaluation of 
the expectation value in \eq\nr{observable2}, 
we start by considering the contribution given by 
${2}\leftrightarrow{2}$ and ${1}\leftrightarrow{3}$ scatterings.
This amounts to a Boltzmann-like computation. 
For Boltzmann equations, matrix elements squared are needed, 
and we first discuss how they can be obtained, 
following ref.~\cite{phasespace}.  

%%%%%%%%%%%%%%%%%%%%%%%%%%% SUBSECTION %%%%%%%%%%%%%%%%%%%%%%%%%%%%%%%%%%%
%
\subsection{Thermal crossing relations for real corrections}
\la{ss:crossing}

Let $\Theta^{ }_{\E}(\P^{ }_1,\P^{ }_2,\P^{ }_3)$ 
denote the matrix element squared
for a $1\to 3$ decay of a sterile neutrino of four-momentum $\K$ into 
final-state particles with momenta $\P^{ }_1$, $\P^{ }_2$ and $\P^{ }_3$.
Here $\E$ defines a generic 
polarization sum, in the sense of \eq\nr{Theta_E}. 
Thermal averaging is denoted by  
\ba
 \scat{1\to3}(a,b,c) & \equiv & 
 \frac{1}{2} \int \! {\rm d}\Omega^{ }_{1\to3} 
          \, \mathcal{N}^{ }_{a,b,c} 
 \;, \la{x_scat1to3a} \\ 
%%%%
 {\rm d}\Omega^{ }_{1\to3} & \equiv & 
 \frac{1}{(2\pi)^9}
 \frac{{\rm d}^3\vec{p}^{ }_a}{2 \epsilon^{ }_{a}}
 \frac{{\rm d}^3\vec{p}^{ }_b}{2 \epsilon^{ }_{b}}
 \frac{{\rm d}^3\vec{p}^{ }_c}{2 \epsilon^{ }_{c}}
 \, (2\pi)^4 \delta^{(4)}(\K
    - \P^{ }_a  - \P^{ }_b - \P^{ }_c  )
 \;,  \\[2mm] 
%%%%
 \mathcal{N}^{ }_{a,b,c} 
 & \equiv & 
 \bar{n}^{ }_{\sigma_a}(\epsilon^{ }_{a} - \mu^{ }_a)
 \,\bar{n}^{ }_{\sigma_b}(\epsilon^{ }_{b} - \mu^{ }_b)
 \,\bar{n}^{ }_{\sigma_c}(\epsilon^{ }_{c} - \mu^{ }_c)
 \nn
 & - &   
   n^{ }_{\sigma_a}(\epsilon^{ }_{a} - \mu^{ }_a)
 \,n^{ }_{\sigma_b}(\epsilon^{ }_{b} - \mu^{ }_b)
 \,n^{ }_{\sigma_c}(\epsilon^{ }_{c} - \mu^{ }_c)
 \;, \la{x_scat1to3c} 
\ea
where we have defined 
\be
 n^{ }_{\sigma}(\epsilon) \; \equiv \; \frac{\sigma}{e^{\epsilon/T} - \sigma}
 \;, \quad
 \bar{n}^{ }_{\sigma}(\epsilon) \; \equiv \; 1 + n^{ }_{\sigma}(\epsilon)
 \;, \la{x_n_sigma}
\ee
in order to permit for a simultaneous treatment of 
Bose-Einstein ($\sigma = +$) and 
Fermi-Dirac distribution functions ($\sigma = -$).
Furthermore on-shell energies are denoted by
\be
 \epsilon^{ }_{a} \; \equiv \; \sqrt{{p}^2_a + m_a^2}
 \;,  \quad
 \omega \; \equiv \; \sqrt{{k}^2 + M^2}
 \;, \la{x_on-shell}
\ee
where 
$ 
 p^{ }_a \equiv |\vec{p}^{ }_a| 
$
and
$ 
 k \equiv |\vec{k}|
$,
so that four-momenta read
$
 \P^{ }_a \; = \; (\epsilon^{ }_{a},\vec{p}^{ }_a)
$
and 
$
 \K \; \equiv \; (\omega,\vec{k}) 
%              \; \equiv \; \K^{ }_2
$.

At the Born level, 
the full rate originating from $2\leftrightarrow 2$ and 
$1\leftrightarrow 3$ reactions can be expressed as 
\ba
 \Gamma^\rmi{Born}_{2\leftrightarrow 2,1\leftrightarrow 3}
 & =  & 
 \bigl[ 
  \scat{1\to3}(a^{ }_1,a^{ }_2,a^{ }_3)
  +  
  \scat{2\to2}(-a^{ }_1;a^{ }_2,a^{ }_3)
  + 
  \scat{2\to2}(-a^{ }_2;a^{ }_3,a^{ }_1)
 \nn[2mm] 
 & + & 
  \scat{2\to2}(-a^{ }_3;a^{ }_1,a^{ }_2)
  + 
  \scat{3\to1}(-a^{ }_1,-a^{ }_2;a^{ }_3)
  + 
  \scat{3\to1}(-a^{ }_3,-a^{ }_1;a^{ }_2)
 \nn[2mm] 
 & + &  
  \scat{3\to1}(-a^{ }_2,-a^{ }_3;a^{ }_1)
  \bigr]
          \, \Theta^{ }_{\E}(\P^{ }_1,\P^{ }_2,\P^{ }_3)
 \;, \la{ex3}
\ea
where the negative index labels indicate that the signs of the 
corresponding momenta are to be inverted. 
Crossed phase space integrals read 
\ba
%%%%%%%%%%%%%%%%%%%%%%%%%%%%%%%%%%%%%%%%%%%
 \scat{2\to2}(-a;b,c) & \equiv & 
 \frac{1}{2} \int \! {\rm d}\Omega^{ }_{2\to2} 
          \, \mathcal{N}^{ }_{a;b,c} 
 \;, \\ 
%%%%
 {\rm d}\Omega^{ }_{2\to2} & \equiv & 
 \frac{1}{(2\pi)^9}
 \frac{{\rm d}^3\vec{p}^{ }_a}{2 \epsilon^{ }_{a}}
 \frac{{\rm d}^3\vec{p}^{ }_b}{2 \epsilon^{ }_{b}}
 \frac{{\rm d}^3\vec{p}^{ }_c}{2 \epsilon^{ }_{c}}
 (2\pi)^4 \delta^{(4)}(\K +  \P^{ }_a 
                       - \P^{ }_b - \P^{ }_c )
 \;,  \\[2mm] 
%%%%
 \mathcal{N}^{ }_{a;b,c} 
 & \equiv & 
 n^{ }_{\sigma_a}(\epsilon^{ }_{a} + \mu^{ }_a)
 \,\bar{n}^{ }_{\sigma_b}(\epsilon^{ }_{b} - \mu^{ }_b)
 \,\bar{n}^{ }_{\sigma_c}(\epsilon^{ }_{c} - \mu^{ }_c)
 \nn 
 & - & 
 \bar{n}^{ }_{\sigma_a}(\epsilon^{ }_{a} + \mu^{ }_a)
 \,n^{ }_{\sigma_b}(\epsilon^{ }_{b} - \mu^{ }_b)
 \,n^{ }_{\sigma_c}(\epsilon^{ }_{c} - \mu^{ }_c)
 \;, \\[2mm]
%%%%%%%%%%%%%%%%%%%%%%%%%%%%%%%%%%%%%%%%%%%%%%%
 \scat{3\to1}(-a,-b;c) & \equiv & 
 \frac{1}{2} \int \! {\rm d}\Omega^{ }_{3\to1} 
          \, \mathcal{N}^{ }_{a,b;c} 
 \;, \\ 
%%%%
 {\rm d}\Omega^{ }_{3\to1} & \equiv & 
 \frac{1}{(2\pi)^9}
 \frac{{\rm d}^3\vec{p}^{ }_a}{2 \epsilon^{ }_{a}}
 \frac{{\rm d}^3\vec{p}^{ }_b}{2 \epsilon^{ }_{b}}
 \frac{{\rm d}^3\vec{p}^{ }_c}{2 \epsilon^{ }_{c}}
 (2\pi)^4 \delta^{(4)}(\K + \P^{ }_a 
                       + \P^{ }_b - \P^{ }_c )
 \;,  \\[2mm] 
%%%%
 \mathcal{N}^{ }_{a,b;c} 
 & \equiv & 
 n^{ }_{\sigma_a}(\epsilon^{ }_{a} + \mu^{ }_a)
 \, n^{ }_{\sigma_b}(\epsilon^{ }_{b} + \mu^{ }_b)
 \, \bar{n}^{ }_{\sigma_c}(\epsilon^{ }_{c} - \mu^{ }_c)
 \nn
 & - & 
 \bar{n}^{ }_{\sigma_a}(\epsilon^{ }_{a} + \mu^{ }_a)
 \,\bar{n}^{ }_{\sigma_b}(\epsilon^{ }_{b} + \mu^{ }_b)
 \,n^{ }_{\sigma_c}(\epsilon^{ }_{c} - \mu^{ }_c)
 \;. \hspace*{6mm}
%%%%%%%%%%%%%%%%%%%%%%%%%%%%%%%%%%%%%%%%%%%%%%%
\ea

The upshot from \eq\nr{ex3} is that 
we only need to compute $\Theta^{ }_{\E}(\P^{ }_1,\P^{ }_2,\P^{ }_3)$
explicitly, with the six other cases obtained ``for free'' from 
crossing symmetries~\cite{phasespace}. It is essential here that
the $1 \to 3$ decays do not need to be kinematically allowed; 
only the analytic structure of $\Theta^{ }_{\E}(\P^{ }_1,\P^{ }_2,\P^{ }_3)$
is needed. 

%%%%%%%%%%%%%%%%%%%%%%%%%%% SUBSECTION %%%%%%%%%%%%%%%%%%%%%%%%%%%%%%%%%%%
%
\subsection{Virtual corrections and mass singularities}
\la{ss:virtuals}

The poles in the matrix element squared, 
$\Theta^{ }_{\E}(\P^{ }_1,\P^{ }_2,\P^{ }_3)$, 
and their residues, 
permit to determine IR-sensitive virtual corrections to 
$1\leftrightarrow 2$ scatterings. 
Generalizing on the KLN theorem~\cite{kln1,kln2}, 
their role is to 
cancel logarithmic and double-logarithmic 
mass singularities from the real 
$2\leftrightarrow 2$ and $1\leftrightarrow 3$
scatterings. 
We do not repeat the rules for this recipe here, 
noting just that a computer-algebraic implementation 
can be found in ref.~\cite{phasespace}. 

%%%%%%%%%%%%%%%%%%%%%%%%%%% SUBSECTION %%%%%%%%%%%%%%%%%%%%%%%%%%%%%%%%%%%
%
\subsection{How to obtain matrix elements squared}

According to \ses\ref{ss:crossing} and \ref{ss:virtuals}, 
the challenge is to extract
the matrix element squared $\Theta^{ }_\E(\P^{ }_1,\P^{ }_2,\P^{ }_3)$
associated with $1\to 3$ decays, with other ingredients 
following from it through established relations. 
In this section we describe how to achieve this. 

%%%%%%%%%%%%%%%%%%%%%%%%%%%%%%% FIGURE %%%%%%%%%%%%%%%%%%%%%%%%%%%%%%%%%%%%%%
%
\begin{figure}[t]
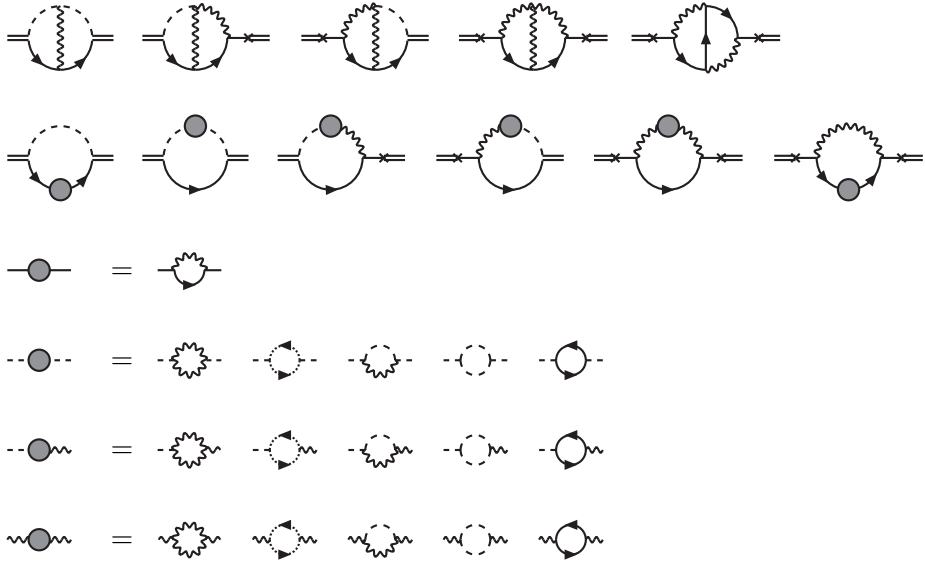


\begin{eqnarray}
&& 
 \hspace*{-1.5cm}
%%%%%%%%%%
 \irredA
 \hspace*{0.00cm}
 \irredB
 \hspace*{0.60cm}
 \irredC
 \hspace*{0.30cm}
 \irredD
 \hspace*{0.50cm}
 \irredE
 \nn[4mm]
&& 
 \hspace*{-1.5cm}
%%%%%%%%%%
 \redA
 \hspace*{0.00cm}
 \redB
 \hspace*{0.00cm}
 \redC
 \hspace*{0.60cm}
 \redD
 \hspace*{0.30cm}
 \redE
 \hspace*{0.60cm}
 \redF
 \nn[3mm] 
%%%%%%%%%%%%%%%%%%%%%%%%%%%%%%%%%%%%
&& 
 \hspace*{-1.5cm}
 \selfFmaster
 \; = \; 
%%%%%%%%%%
 \selfFA
 \nn
%%%%%%%%%%%%%%%%%%%%%%%%%%%%%%%%%%%%%%%%%%%%%%%%%%%%
&& 
 \hspace*{-1.5cm}
 \selfSmaster
 \; = \;
%%%%%%%%%%
 \selfSA
 \hspace*{0.00cm}
 \selfSB
 \hspace*{0.00cm}
 \selfSF
 \hspace*{0.00cm}
 \selfSD
 \hspace*{0.00cm}
 \selfSG
 \nn 
%%%%%%%%%%%%%%%%%%%%%%%%%%%%%%%%%%%%%%%%%%%%%%%%%%%%
&& 
 \hspace*{-1.5cm}
 \selfMmaster
 \; = \;
%%%%%%%%%%
 \selfMA
 \hspace*{0.00cm}
 \selfMB
 \hspace*{0.00cm}
 \selfMF
 \hspace*{0.00cm}
 \selfMD
 \hspace*{0.00cm}
 \selfMG
 \nn 
%%%%%%%%%%%%%%%%%%%%%%%%%%%%%%%%%%%%%%%%%%%%%%%%%%%%
&& 
 \hspace*{-1.5cm}
 \selfGmaster
 \; = \;
%%%%%%%%%%
 \selfA
 \hspace*{0.00cm}
 \selfB
 \hspace*{0.00cm}
 \selfF
 \hspace*{0.00cm}
 \selfD
 \hspace*{0.00cm}
 \selfG
 \nonumber
\end{eqnarray}

\vspace*{-4mm}

\caption[a]{\small 
 2-loop self-energy diagrams possessing a non-vanishing cut, 
 corresponding to 
 $2\leftrightarrow2$ and $1\leftrightarrow3$ scatterings. 
 Double lines correspond to sterile neutrinos; 
 solid lines to fermions; 
 dashed lines to scalars; 
 wiggly lines to gauge fields; 
 crosses to a mixing between sterile and active neutrinos, 
 induced by a Higgs vev. 
 Lepton Yukawa couplings have been omitted for simplicity. 
} 
\la{fig:retarded}
\end{figure}
%
%%%%%%%%%%%%%%%%%%%%%%%%%%%%%%%%%%%%%%%%%%%%%%%%%%%%%%%%%%%%%%%%%%%%%%%%%%%%%

One way to obtain $\Theta^{ }_{\E}$ 
is to determine the retarded correlator of \eq\nr{PiR}, 
and then to extract its cut. Viewing the operators as in 
\eq\nr{observable2}, the Feynman
diagrams contributing to the retarded correlator are those
in \fig\ref{fig:retarded}. 

This type of computations are conveniently implemented with FORM \cite{form}.
There are, however, challenges. The result obtained after
Wick contractions contains many ambiguities, related to choices of 
sum-integration variables. Even just to demonstrate the gauge 
independence of the retarded correlator, 
we need to make use of 
renamings of variables and of integration-by-parts identities,
in order to eliminate the ambiguities. 
Even if this can be done and gauge independence verified,  
there is a price to pay for these reductions, namely that 
when the cut is extracted, the remaining 
integrands do not display their natural symmetries, 
as all symmetries have been eliminated
in favour of an unambiguous representation.  

In order to illustrate the situation, let us define ``master'' 
sum-integrals, general enough such that all 2-loop contributions can 
be determined as their linear combinations:
\ba
 && \hspace*{-1.5cm}
 I^{j_1\cdots j_6 }_{i_1\cdots i_6}(a^{ }_1,\ldots,a^{ }_6) 
 \nn 
 & \equiv & 
   \Tint{P,Q} \!\! 
   \frac{j^{ }_1 H_{i\tilde P}
        +j^{ }_2 H_{i\tilde Q}
        +j^{ }_3 H_{i(\tilde P-\tilde Q)} 
        +j^{ }_4 H_{-i(\tilde K+\tilde P)}
        +j^{ }_5 H_{-i(\tilde K+\tilde Q)}
        +j^{ }_6 H_{-i\tilde K} }
   { 
     \Delta^{i_1}_{P;a^{ }_1}
     \,\Delta^{i_2}_{Q;a^{ }_2}
     \,\Delta^{i_3}_{P-Q;a^{ }_3}
     \,\Delta^{i_4}_{-K-P;a^{ }_4}
     \,\Delta^{i_5}_{-K-Q;a^{ }_5}
     \,\Delta^{i_6}_{-K;a^{ }_6}
   }
 \;. \la{master} 
\ea
Here $\Tinti{P} \equiv T \sum_{p_n} \int_\vec{p}$ is a Matsubara 
sum-integral, with $p^{ }_n$ referring either to a bosonic or 
fermionic Matsubara frequency, and $P \equiv (p^{ }_n,\vec{p})$.
The inverse propagators read 
\be
 \Delta^{ }_{P;a_x} 
 \; \equiv \; (p^{ }_n + i \mu^{ }_{a_x})^2 + p^2 + m_{a_x}^2
 \;, \la{prop}
\ee
where $\mu^{ }_{a_x}$ is the chemical potential associated
with the particle species ${a^{ }_x}$, and $m^{ }_{a_x}$ is its mass,  
whereas $H^{ }_{i \tilde P}$ denotes a helicity projection, 
for instance 
$ 
 H^{ }_{i \tilde P} = \E \cdot i \tilde P
$
for our problem. 

In this language, symmetries depend on how many of 
the indices $i^{ }_1,...,i^{ }_5$ are positive and whether some of the
corresponding propagators carry identical particles.   
For instance, if $i^{ }_4 \le 0$ and $a^{ }_1 = a^{ }_3$, we can 
make use of the substitution $P\to Q-P$ to interchange the 1st and
3rd propagator, replacing the original master by $1/2$ times the
sum of the original and reflected master.  
However, the momenta come with opposite signs
after the substitution, so this is a symmetry only if the particle
in question carries no chemical potential. Furthermore the numerator
structures weighted by $j^{ }_1$, $j^{ }_3$ and $j^{ }_4$ change, 
and need to be projected back to the basis. Finally, if 
$i^{ }_4 < 0$, this inverse propagator needs to be expressed 
in terms of the new inverse propagators, via
\ba
   \Delta^{ }_{-K-Q+P;a_4} 
   & = & 
   \Delta^{ }_{P-Q;-a_1}
 - \Delta^{ }_{Q;a_2}
 + \Delta^{ }_{P;-a_3}
 - \Delta^{ }_{-K-P;a_4} 
 + \Delta^{ }_{-K-Q;a_5}
 + \Delta^{ }_{-K;a_6}
 \nn 
 & - &  
 m_{a_1}^2 + m_{a_2}^2 - m_{a_3}^2 + 2 m_{a_4}^2 - m_{a_5}^2 - m_{a_6}^2
 \;. 
\ea
This assumes that the chemical potentials of the different species
are related to each other in a way guaranteeing chemical equilibrium
(cf.\ the paragraph following \eq\nr{tildeP}). 
Now, if the numerator structures are invariant in the reflection, 
the appearance of a minus sign in front of  
$ \Delta^{ }_{-K-P;a_4} $ 
on the right-hand side ensures 
that this part is eliminated 
when the original and reflected term are averaged over. 
Had we not considered this symmetrization, 
a redundant term, with a possibly
gauge-dependent coefficient,  
would have remained. 

The most extensive symmetrizations appear when there are just
three propagators, which is the minimal case leading to 
$2\leftrightarrow2$ and $1\leftrightarrow3$ cuts. Prime examples 
are $I^{j_1 \cdots j_6}_{1 i_2 1 i_4 1 i_6}$, with $i^{ }_2,i^{ }_4 \le 0$,  
and $I^{j_1 \cdots j_6}_{i_1 1 1 1\, i_5 i_6}$, with $i^{ }_1,i^{ }_5 \le 0$.
In these cases, $3!  =6$ momentum permutations need to 
be implemented, in order to eliminate ambiguities and 
establish gauge independence.  

As a bottom line for this approach, we note that it is ideal if  
a minimal set of master integrals are evaluated afterwards. If, however, 
a nice-looking intermediate expression is needed, such as our 
$\Theta^{ }_{\E}(\P^{ }_1,\P^{ }_2,\P^{ }_3)$, 
which should display maximal symmetries, 
then the elimination of all redundancies is 
somewhat counterproductive. 

%%%%%%%%%%%%%%%%%%%%%%%%%%%%%%% FIGURE %%%%%%%%%%%%%%%%%%%%%%%%%%%%%%%%%%%%%%
%
\begin{figure}[t]
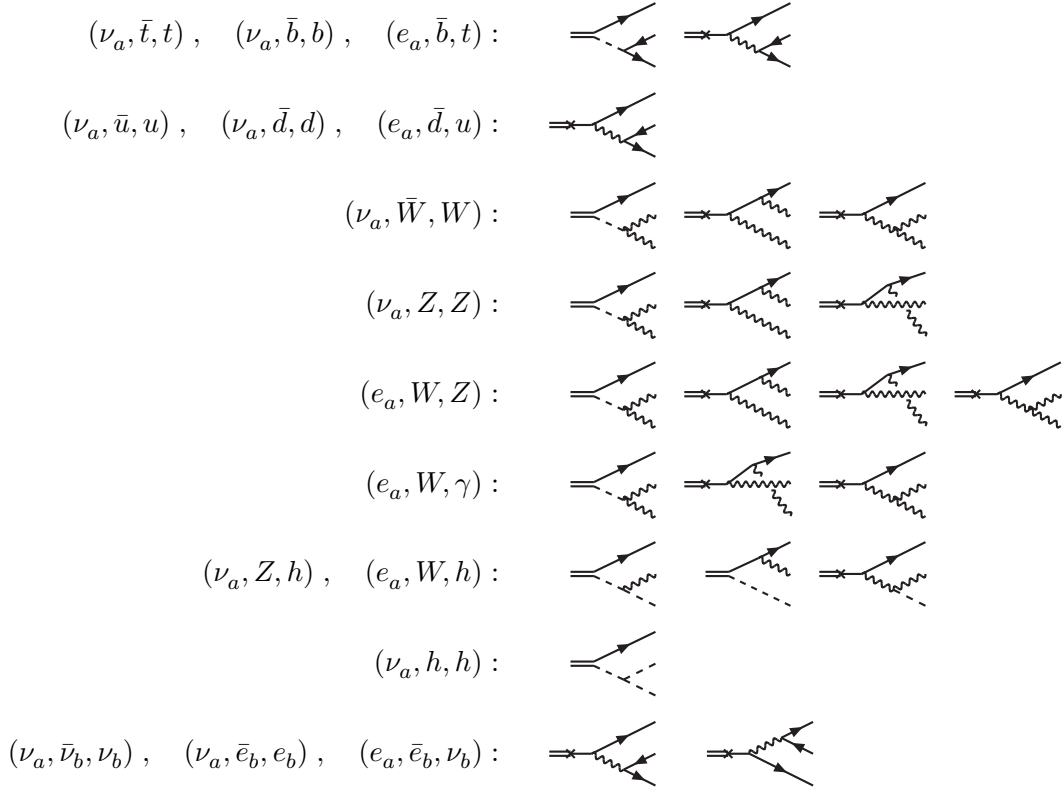


%\hspace*{1.5cm}%
%\begin{minipage}[c]{3cm}
\begin{eqnarray*}
 (\ana, \Atop,\atop)\;, \quad
 (\ana, \Abot,\abot)\;, \quad
 (\aea, \Abot,\atop): 
&& 
 \decB
 \hspace*{-0.00cm}
 \decA
 \hspace*{-0.00cm}
 \nn 
%%%%%%%%%%
 (\ana, \Aup,\aup)\;, \quad
 (\ana, \Ado,\ado)\;, \quad
 (\aea, \Ado,\aup): 
&& 
 \decA
 \hspace*{-0.00cm}
 \nn 
%%%%%%%%%%
 (\ana,\AW,\aW): 
&& 
 \decI
 \hspace*{-0.00cm}
 \decG
 \hspace*{-0.00cm}
 \decJ
 \hspace*{-0.00cm}
 \nn 
%%%%%%%%%%
 (\ana,\aZ,\aZ): 
&& 
 \decI
 \hspace*{-0.00cm}
 \decG
 \hspace*{-0.00cm}
 \decH
 \hspace*{-0.00cm}
 \nn 
%%%%%%%%%%
 (\aea,\aW,\aZ): 
&& 
 \decI
 \hspace*{-0.00cm}
 \decG
 \hspace*{-0.00cm}
 \decH
 \hspace*{-0.00cm}
 \decJ
 \hspace*{-0.00cm}
 \nn 
%%%%%%%%%%
 (\aea,\aW,\aQ): 
&& 
 \decI
 \hspace*{-0.00cm}
 \decH
 \hspace*{-0.00cm}
 \decJ
 \hspace*{-0.00cm}
 \nn 
%%%%%%%%%%
 (\ana,\aZ,\ah)\;, \quad
 (\aea,\aW,\ah): 
&& 
 \decE
 \hspace*{-0.00cm}
 \decF
 \hspace*{-0.00cm}
 \decD
 \hspace*{-0.00cm}
 \nn 
%%%%%%%%%%
 (\ana,\ah,\ah): 
&& 
 \decK
 \hspace*{-0.00cm}
 \nn 
%%%%%%%%%%
% (\ana, \Ana,\ana)\;, \quad
% (\ana, \Aea,\aea): 
% && 
 (\ana, \An,\an)\;, \quad
 (\ana, \Ae,\aeb)\;, \quad
 (\aea, \Ae,\an): 
&& 
 \decA
 \hspace*{0.30cm}
 \decC
 \hspace*{-0.00cm}
 \nn 
%%%%%%%%%%
% (\ana, \An,\an)\;, \quad
% (\ana, \Ae,\aeb)\;, \quad
% (\aea, \Ae,\an): 
% && 
% \decA
% \hspace*{-0.00cm}
% \nn 
%%%%%%%%%%
\end{eqnarray*}
%\end{minipage}

\vspace*{-4mm}

\caption[a]{\small 
 $1\rightarrow 3$ contributions to  
 sterile neutrino interaction rate. The notation is as 
 in \fig\ref{fig:retarded}, with in addition 
 $\aW \,\equiv\, W^+_{ }$ and  
 $\AW \,\equiv\, W^-_{ }$.  
 The labelling of the decay products goes from top to bottom. 
 The $2\leftrightarrow 2$ and $3\rightarrow 1$ 
 contributions can be obtained by pulling one or two
 legs to the left, respectively. 
} 
\la{fig:1to3}
\end{figure}
%
%%%%%%%%%%%%%%%%%%%%%%%%%%%%%%%%%%%%%%%%%%%%%%%%%%%%%%%%%%%%%%%%%%%%%%%%%%%%%

Another strategy is to determine directly the matrix elements
squared appearing in the decay rate, by considering the amplitudes
depicted in \fig\ref{fig:1to3}. For this kind of computations, many
automated tools are available, though we find it simplest to again
implement the algebra with FORM~\cite{form}.  

As a remark on the latter approach, we note that
even if it works well for simple amplitudes, 
the more complicated cases, with many interference terms, 
are not identified in a nice form. 
Indeed, as kinematic variables are not independent, several equivalent 
representations can be given to any matrix element squared. 
In general, human inspection is required for judging 
which variables yield the ``nicest'' expression. 
Our results after some undoubtedly incomplete efforts
in this direction are collected in appendix~A.

%%%%%%%%%%%%%%%%%%%%%%%%%%% SUBSECTION %%%%%%%%%%%%%%%%%%%%%%%%%%%%%%%%%%%
%
\subsection{Splitup into direct and indirect contributions}
\la{ss:splitup_2to2}

As indicated by \eq\nr{observable2} and illustrated 
by \figs\ref{fig:retarded} and \ref{fig:1to3}, 
the result for 
% $2\leftrightarrow 2$ and $1\leftrightarrow 3$ scatterings 
$1\to 3$ decays does {\em not} get trivially 
separated into indirect and direct contributions, but there 
are interference terms between these two sets. 
Nevertheless, such a splitup is still possible, 
as we now show.\footnote{%
 The representation is not unique but it can be found. 
 }  

The idea is to first compute all the contributions 
in \fig\ref{fig:retarded} or \ref{fig:1to3}. 
At this point, we have verified
the gauge independence of the full expression. Then, we view the 
result as an expansion in the virtuality of the sterile neutrinos, 
$\MM$. The expansion contains two singular orders, 
$1/M^4$ and $1/\MM$. It turns out that the coefficients of 
these terms are not independent, but are related to each other
in a specific way. This permits for us to ``resum'' these
singular terms into an indirect contribution. 

More concretely, let us define 
\be
 \Proj \; \equiv \; \frac{\E}{M^2} - \frac{2\E\cdot\K\, \K}{M^4}
 \;, \la{Proj}
\ee
where $\mathcal{E}$ is the external four-vector from \eq\nr{Theta_E}. 
As demonstrated by the explicit expressions in appendix~\ref{app:1to3}, 
all appearances of inverse powers of $\MM$ can be represented through
scalar products containing $\Proj$.
Subsequently, we can write\hspace*{0.1mm}\footnote{%
 To clarify the appearance of the chiral projectors, we recall that 
 if a matrix effectuates a transition from the left to the right 
 subspace, 
 $ 
   w^{ }_\rmii{R} = M v^{ }_\rmii{L}
 $,  
 then its inverse operates in the opposite direction, 
 $ 
   v^{ }_\rmii{L} = M^{-1}_{ } w^{ }_\rmii{R}
 $.
 }
\ba
 \Proj\cdot\mathcal{S}  & \subset & 
 \frac{v^2}{2} \, \im \tr \Bigl\{ \msl{\mathcal{E}} 
 \aL\, \Bigl[\aR \bigl( {- \bsl{\K} - \bsl{\Sigma}} \bigr) \aL \Bigr]^{-1}
 \; \aR  \Bigr\} 
 \;, \la{resum0} \\
 \im \Sigma & \supset & - \, \frac{\mathcal{S}}{v^2}
 \;. \la{resum}
\ea
The self-energy $\Sigma$ represents the indirect contribution.
The parts of 
$\Theta^{ }_{\E}$ containing scalar products with $\E$ 
(amounting to direct contributions)
and $\Proj$ 
(amounting to indirect contributions)
are listed in 
\eqs\nr{hadronic_broken} and \nr{leptonic_broken} for hadronic 
and leptonic effects, respectively. 

%%%%%%%%%%%%%%%%%%%%%%%%%%% SUBSECTION %%%%%%%%%%%%%%%%%%%%%%%%%%%%%%%%%%%
%
\subsection{HTL resummation}
\la{ss:HTL}

If the temperature is large compared with all masses 
($m^{ }_i \ll \pi T$), then the thermal masses generated
by Standard Model interactions ($\delta^{ }_\T m^{ }_i \sim gT$) can be
as large as the vacuum masses, and needed to be included for a consistent
computation. The systematic way to do this goes through
Hard Thermal Loop resummation~\cite{ht1,ht2,ht3,ht4}.

Considering first either the side of the symmetric phase
(cf.\ \eqs\nr{hadronic_symmetric} and \nr{leptonic_symmetric}), 
or direct contributions in the Higgs phase, 
i.e.\ effects proportional to $\E$ in 
\eqs\nr{hadronic_broken} and \nr{leptonic_broken},
the terms are integrable after regularizing 
lepton propagators by a small mass~\cite{phasespace}.
As reviewed in \se{5} of ref.~\cite{phasespace}, the regulator
could subsequently be replaced by a physical thermal lepton 
mass, via an addition-subtraction step analogous to that 
in \eq\nr{Sigma_resummed}. However, if $M\sim \pi T$, 
this is not necessary at leading order, because the 
large energy in $M$ guarantees that the soft phase space
region gives a subdominant contribution.\footnote{%  
  In some older computations, the Higgs mass $m^{ }_{\aS}$ had
  been set to zero in $2\leftrightarrow 2$ scatterings. Then, 
  following \se{6.1} of ref.~\cite{degenerate}, 
  a similar addition-subtraction treatment
  is needed for the Higgs mass as well. 
  }

Turning to the indirect contribution, 
it can be anticipated from \eq\nr{resum} that 
$\im\Sigma$ can be singular if $v\to 0$, i.e.\ if masses are sent 
to zero. Put another way, the division by $v^2$ implies that masses
$\sim g^2 v^2$ turn into couplings $\sim g^2$. 
Therefore, as can also be deduced from \fig\ref{fig:1to3}, 
$\omega \im\Sigma$ is of $\rmO(g^4)$ 
in terms of couplings originating from vertices. At the same 
time, the phase space average is quadratically divergent at small
momenta if masses are sent to zero~\cite{broken}. 
Given that the $v$-independent scales 
characterizing the thermal average are $\sim \pi T$ and $M$, 
the divergence starts to play a role if $v \ll \max \{ M,\pi T\}$.   

If we are in the regime $M \ll \pi T$, 
the would-be divergence can be cured 
by HTL resummation, which gives the 
gauge bosons thermal masses $\sim g T$. The thermal masses
``regulate'' the propagators in the regime $g v \ll g T$.
For the problem at hand, leading-order HTL resummation 
has been worked out in \se{5.2} of ref.~\cite{degenerate},  
and the NLO level has been reached in ref.~\cite{nlo_width}. 
The magnitude of $\omega \im\Sigma$ is $\rmO(g^2 T^2)$
after the resummation. 

In a practical computation,  
we normally need to {\em interpolate}
between domains in which HTL resummation is important or not. 
In order to avoid double counting, HTL resummation  
must then be implemented through an addition-subtraction step, 
\be
 \bsl{\Sigma}^\rmi{resummed}_{ }
 = 
 \bsl{\Sigma}^\rmi{Born}_{ }
 + 
% \Lambda^{ }_{ }\, \bigl( \, 
 \bsl{\Sigma}^\rmi{HTL,full}_{ }
 - 
 \bsl{\Sigma}^\rmi{HTL,expanded}_{ }
% \, \bigr) 
 \;, \la{Sigma_resummed}
\ee 
where the subtraction removes terms that would be part both of 
$
 \bsl{\Sigma}^\rmi{Born}_{ }
$
and
$
 \bsl{\Sigma}^\rmi{HTL,full}_{ }
$.
Let us stress again, however, that if $M \gsim \pi T$, HTL resummation
is not necessary at leading order for the observables that we are
concerned with. 

%%%%%%%%%%%%%%%%%%%%%%%%%%% SECTION %%%%%%%%%%%%%%%%%%%%%%%%%%%%%%%%%%%%%%
%
\section{Rates from 
${1+n}\leftrightarrow{2+n}$ scatterings}
\la{se:sterile_1to2}

Finally we consider sterile neutrino rate coefficients originating
from ${1+n}\leftrightarrow{2+n}$ reactions. Again we obtain both 
direct and indirect contributions. Following ref.~\cite{lpm}, we 
carry out the basic discussion on the level of $1\leftrightarrow 2$
reactions (cf.\ \se\ref{ss:splitup_1to2}), 
noting that LPM resummation over $n\ge 0$ 
(cf.\ \se\ref{ss:LPM})
can be implemented through a subsequent addition-subtraction step, 
analogous to \eq\nr{Sigma_resummed}. 

%%%%%%%%%%%%%%%%%%%%%%%%%%% SUBSECTION %%%%%%%%%%%%%%%%%%%%%%%%%%%%%%%%%%%
%
\subsection{Splitup into direct and indirect contributions}
\la{ss:splitup_1to2}

%%%%%%%%%%%%%%%%%%%%%%%%%%%%%%% FIGURE %%%%%%%%%%%%%%%%%%%%%%%%%%%%%%%%%%%%%%
%
\begin{figure}[t]
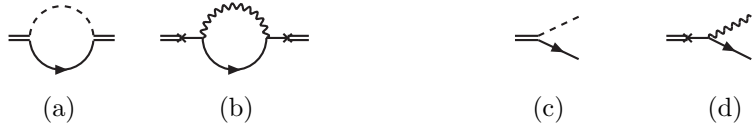


\begin{eqnarray}
&& 
 \hspace*{-0.0cm}
%%%%%%%%%%
 \redX
 \hspace*{0.50cm}
 \redY
 \hspace*{2.50cm}
 \decBB
 \hspace*{0.50cm}
 \decAA
 \nn[1mm] 
&&
 \hspace*{0.55cm} \mbox{\small (a)}
 \hspace*{1.9cm} \mbox{\small (b)}
 \hspace*{3.7cm} \mbox{\small (c)}
 \hspace*{1.9cm} \mbox{\small (d)}
 \nonumber
\end{eqnarray}

\vspace*{-4mm}

\caption[a]{\small 
 Left: 1-loop self-energy diagrams possessing a non-vanishing cut, 
 with the same notation as in \fig\ref{fig:retarded}.
 Right: the corresponding amplitudes,
 with the same notation
 as in \fig\ref{fig:1to3}. 
} 
\la{fig:1to2}
\end{figure}
%
%%%%%%%%%%%%%%%%%%%%%%%%%%%%%%%%%%%%%%%%%%%%%%%%%%%%%%%%%%%%%%%%%%%%%%%%%%%%%

To set the stage, we start by considering 
Feynman diagrams in the Higgs phase, 
shown in \fig\ref{fig:1to2} 
(the diagrams on the left are analogous to the diagrams
in \fig\ref{fig:retarded} but at 1-loop level).
As was the case with the 2-loop diagrams 
in \fig\ref{fig:retarded}, 
the two contributions are not gauge-independent by themselves.
Thus, gauge independence needs to be verified first, 
before splitting the contributions into the direct and indirect ones. 
Let us describe this procedure, which is similar to that introduced 
around \eq\nr{resum}, in some more detail.\footnote{%
 Elaborations on the gauge dependence 
 of $1\leftrightarrow 2$ processes 
 can also be found in ref.~\cite{1to2}.
 } 

Omitting tadpole-like terms that are related to the
renormalization of the tree-level contribution from the first
row of \eq\nr{observable2}, 
the diagrams in \fig\ref{fig:1to2} amount to 
\ba
 \Pi^\rmii{E}_a |^{ }_\rmi{(a)}  & = &  \frac{1}{2}\, 
 \aL\, \bsl{\Delta}^{-1}_{P;\nu^{ }_a} \aR 
 \, \bigl[ \Delta^{-1}_{-K-P;h} + \Delta^{-1}_{-K-P;GZ} \bigr]
 + 
 \aL\, \bsl{\Delta}^{-1}_{P;e^{ }_a} \aR 
 \, \Delta^{-1}_{-K-P;GW}
 \;, \la{1to2_dir} \\ 
%%%%%%%
 \Pi^\rmii{E}_a |^{ }_\rmi{(b)}  & \supset &  \frac{1}{2}\, 
 \aL\, \bsl{\Delta}^{-1}_{P;\nu^{ }_a} \aR 
 \, \bigl[ \Delta^{-1}_{-K-P;Z} - \Delta^{-1}_{-K-P;Z'} \bigr]
% \nn 
% & + &  
 + 
 \aL\, \bsl{\Delta}^{-1}_{P;e^{ }_a} \aR 
 \, \bigl[ \Delta^{-1}_{-K-P;W} - \Delta^{-1}_{-K-P;W'} \bigr]
 \nn 
 & + & \mZ^2\, \aL\, 
             \biggl(  \frac{\bsl{\Delta}^{-1}_{P;\nu^{ }_a}}{\tilde K^2}
               + \frac{2 \tilde K\cdot i\tilde P 
                      \, \bsl{\tilde K} \Delta^{-1}_{P;\nu^{ }_a}}{\tilde K^4}
             \biggr)
             \,\aR \Delta^{-1}_{-K-P;Z} 
 \nn 
 & + & 2\mW^2\, \aL\, 
             \biggl(  \frac{\bsl{\Delta}^{-1}_{P;e^{ }_a}}{\tilde K^2}
               + \frac{2 \tilde K\cdot i\tilde P 
                      \, \bsl{\tilde K} \Delta^{-1}_{P;e^{ }_a}}{\tilde K^4}
             \biggr)
             \,\aR \Delta^{-1}_{-K-P;W} 
 \;, \la{1to2_indir}
\ea
where 
$\bsl{\Delta}^{-1}_{ }$ and 
$\Delta^{-1}$ are fermion and scalar
propagators, respectively; 
a sum-integral over $P$ is implied; 
$GZ$, $GW$ stand for neutral and charged Goldstone modes; 
and $Z',W'$ refer to 
the longitudinal gauge modes, with 
$\mZp^2 \equiv \xi \mZ^2$ and 
$\mWp^2 \equiv \xi \mW^2$ 
where $\xi$ is a gauge parameter.  

It is obvious from \eqs\nr{1to2_dir} and \nr{1to2_indir} that the two parts
are not gauge independent separately, as the former depends on the Goldstone
masses and the latter on $\mZp^2$ and $\mWp^2$. 
Adding together and noting that in the 
tree-level vacuum, 
$ \mGZ^2  = \mZp^2$ and 
$ \mGW^2  = \mWp^2$, 
gauge dependence drops out, 
because of opposite signs on the first rows. 

How the cancellation operates
is that the first row of \eq\nr{1to2_indir}  effectively
replaces the Goldstone masses in \eq\nr{1to2_dir} through
physical gauge boson masses. The same result could be obtained by 
evaluating \eq\nr{1to2_dir} in Feynman gauge, as in that case 
$\mGZ^{ } = \mZ^{ }$  and 
$\mGW^{ } = \mW^{ }$. 
We may call this the direct contribution. 

The remaining terms in \eq\nr{1to2_indir} are proportional to inverse
powers of $\tilde{K}^2$. 
They become large if we analytically continue
$\tilde{K}^2 \to - M^2$
and then make $M^2$ small. These terms 
need to be ``resummed'' into an indirect contribution, 
in analogy with \eqs\nr{resum0} and \nr{resum}. In total, the  
$1\leftrightarrow 2$ processes can thus be represented as 
\ba
 \Pi^\rmii{E}_a |^{ }_\rmi{direct}  & = &  \frac{1}{2} 
 \,\aL\, \bsl{\Delta}^{-1}_{P;\nu^{ }_a} \aR 
 \, \bigl[ \Delta^{-1}_{-K-P;h} + \Delta^{-1}_{-K-P;Z} \bigr]
 + 
 \aL\, \bsl{\Delta}^{-1}_{P;e^{ }_a} \aR 
 \, \Delta^{-1}_{-K-P;W}
 \;, \la{1to2_dir_final} \\ 
%%%%%%%
 \Pi^\rmii{E}_a |^{ }_\rmi{indirect}  & = &  
 \frac{v^2}{2} \, 
 \aL \bigl[ 
 \aR \bigl( {-i \bsl{\tilde K} - \bsl{\Sigma}} \bigr) \aL \bigr]^{-1} \aR
 \;, \la{1to2_indir_strctr} \\ 
%%%%%%%
 \aR \bsl{\Sigma} \aL  & = & \frac{g_1^2 + g_2^2}{2}
  \, \aR\, \bsl{\Delta}^{-1}_{P;\nu^{ }_a} 
             \aL \Delta^{-1}_{-K-P;Z} 
 + g_2^2 \, \aR\, \bsl{\Delta}^{-1}_{P;e^{ }_a}
             \aL \Delta^{-1}_{-K-P;W} 
 \;. \la{1to2_indir_final}
\ea
The zeroth order term in \eq\nr{1to2_indir_strctr} 
originates from a tree-level evaluation of 
the first row of \eq\nr{observable2}. 
If we go to the symmetric phase, i.e.\ $v\to 0$, 
then only the direct contribution is present. 
That result is given by \eq\nr{1to2_dir_final}, with 
the substitutions $\nu^{ }_a,e^{ }_a \to \ell^{ }_a$ and $h,Z,W \to \aS$.

%%%%%%%%%%%%%%%%%%%%%%%%%%% SUBSECTION %%%%%%%%%%%%%%%%%%%%%%%%%%%%%%%%%%%
%
\subsection{LPM resummation}
\la{ss:LPM}

If the particles participating in $1\leftrightarrow 2$ scatterings
are ultrarelativistic ($m^{ }_i \ll \epsilon^{ }_i$, $M \ll \omega$)
and if their virtualities are small compared with the rate of 
thermal scatterings ($m_i^2/\epsilon^{ }_i, M^2/\omega \ll g^2 T/\pi$), 
then many scatterings of the type $1+n \leftrightarrow 2 +n$ take 
place, and need to be summed to all orders, through so-called 
LPM resummation~\cite{gelis3,amy1,agmz,bb1}. 
An implementation of LPM resummation guaranteeing that 
the Born limits of \se\ref{ss:splitup_1to2} are correctly recovered 
when we exit the ultrarelativistic regime, 
has been worked out in ref.~\cite{lpm}. 
However, if $M$ is not small compared with thermal scales, 
for instance if we find ourselves in the regime 
$M\sim \pi T$, then no LPM resummation is needed. 

%%%%%%%%%%%%%%%%%%%%%%%%%%% SECTION %%%%%%%%%%%%%%%%%%%%%%%%%%%%%%%%%%%%%%
\section{Mass corrections}
\la{se:mass}

%%%%%%%%%%%%%%%%%%%%%%%%%%% SUBSECTION %%%%%%%%%%%%%%%%%%%%%%%%%%%%%%%%%%%
%
\subsection{Real part of the direct contribution}
\la{ss:dir_real}

The retarded correlator of \eq\nr{PiR} contains also a real part,
yielding dispersive corrections according to \eq\nr{def_U}. 
In contrast to the imaginary part, where a ``cut'' leads to 
phase-space constraints and separate classes
of diagrams (cf.\ \ses\ref{se:sterile_2to2} and 
\ref{se:sterile_1to2}), the real part contains 
an unconstrained integral, whose magnitude is simpler to estimate. 
Normally, it is sufficient to restrict to 1-loop level in its evaluation. 
Then the direct contribution originates from 
\eq\nr{1to2_dir_final} and the indirect one from
\eq\nr{1to2_indir_final}.\footnote{%
 The latter includes the familiar tree-level 
 correction of $\rmO(h_\nu^2 v^2 / M)$, 
 once neutrino Yukawas are included.
 } 

To be concrete, let us define a master sum-integral 
\be
 J^{j_1 j_2 j_3}_{i_1 i_2 i_3}(a^{ }_1,a^{ }_2,a^{ }_3) 
 \; \equiv \; 
 \Tint{P} 
    \frac{            j^{ }_1 \{ i \bsl{ \tilde P }\! \}
                    + j^{ }_2 \{-i( \bsl{ \tilde K }\!
                                  + \bsl{ \tilde P }\! ) \}
                    + j^{ }_3 \{-i \bsl{ \tilde K }\! \} 
   }
   { 
     \Delta^{i_1}_{P;a^{ }_1}
     \,\Delta^{i_2}_{-K-P;a^{ }_2}
     \,\Delta^{i_3}_{-K;a^{ }_3}
   }
 \;, \la{master1}
\ee
where the notation adheres to that in \eq\nr{master}. 
Then we may rewrite \eq\nr{1to2_dir_final} as
\be
 \Pi^\rmii{E}_a |^{ }_\rmi{direct} =  
 \aL \, \biggl\{ 
        J^{ -\frac{1}{2} 0 0}_{1\, 1\, 0}
        ( \ana,\ah, 0 )
 + 
        J^{ -\frac{1}{2} 0 0}_{1\, 1\, 0}
        ( \ana,\aZ, 0 )
 + 
 2  \,
        J^{ -\frac{1}{2} 0 0}_{1\, 1\, 0}
        ( \aea,\aW, 0 )
 \biggr\}\, \aR
 \;. \la{dir_1to2_general}
\ee  

The real part is, in general, ultraviolet divergent. As the divergences
are related to NLO corrections, notably wave function renormalization, 
we address only the part proportional
to the plasma four-velocity, which is finite and plays a qualitatively
more prominent role. Following ref.~\cite{degenerate}, 
this part is defined by writing 
\be
  \re J^{ -\frac{1}{2} 0 0}_{1\, 1\, 0}
        ( {a}^{ }_1,{a}^{ }_2, 0 ) 
        \bigr|^{ }_{\tilde k_n \to -i [\omega + i0^+]}
  \; \equiv \; 
  \mathcal{A} \, \bsl{\K} + 
  \V^{ -\frac{1}{2} 0 0}_{1\, 1\, 0}
        ( {a}^{ }_1,{a}^{ }_2, 0 ) \, \msl{\U}
 \;. \la{V_def}
\ee
By projecting onto the part proportional to~$\msl{\U}$, 
a straightforward analysis leads to 
\ba
 & & \hspace*{-1.3cm} 8 \pi^2 k^3 
  \V^{ -\frac{1}{2} 0 0}_{1\, 1\, 0}
        ( {a}^{ }_1,{a}^{ }_2, 0 ) 
%%% 
 \la{re_J_result} \\[2mm] 
 & = & 
 \int_{m^{ }_1}^{\infty} \! {\rm d} \epsilon^{ }_1
 \, \biggl\{ 
  n^{ }_{\sigma_1}(\epsilon^{ }_1 + \mu^{ }_1) \, 
  \biggl[
     \frac{\omega k p^{ }_1}{2}
%%% 
\nn 
  & + &   
    \frac{\omega(m_2^2 - m_1^2 - M^2) - 2 \epsilon^{ }_1 M^2}{8} 
    \ln \biggl| 
      \frac{m_2^2 - m_1^2 - M^2 - 2 (k p^{ }_1 + \omega \epsilon^{ }_1)}
           {m_2^2 - m_1^2 - M^2 + 2 (k p^{ }_1 - \omega \epsilon^{ }_1)}
    \biggr|
  \biggr]
%%% 
\nn 
 & + & 
  n^{ }_{\sigma_1}(\epsilon^{ }_1 - \mu^{ }_1) \, 
  \biggl[
     \frac{\omega k p^{ }_1}{2}
%%% 
\nn 
  & + &   
    \frac{\omega(m_2^2 - m_1^2 - M^2) + 2 \epsilon^{ }_1 M^2}{8} 
    \ln \biggl| 
      \frac{m_2^2 - m_1^2 - M^2 - 2 (k p^{ }_1 - \omega \epsilon^{ }_1)}
           {m_2^2 - m_1^2 - M^2 + 2 (k p^{ }_1 + \omega \epsilon^{ }_1)}
    \biggr|
  \biggr]
%%%%% 
 \biggr\}^{ }_{p^{ }_1 \equiv \sqrt{\epsilon_1^2 - m_1^2}}
%%%%%%%%%%%%%%
 \nn 
 & - & 
 \int_{m^{ }_2}^{\infty} \! {\rm d} \epsilon^{ }_2
 \, \biggl\{ 
  n^{ }_{\sigma_2}(\epsilon^{ }_2 + \mu^{ }_2) \, 
  \biggl[
     \frac{\omega k p^{ }_2}{2}
%%% 
\nn 
  & + &   
    \frac{\omega(m_1^2 - m_2^2 - M^2) - 2 \epsilon^{ }_2 M^2}{8} 
    \ln \biggl| 
      \frac{m_1^2 - m_2^2 - M^2 - 2 (k p^{ }_2 + \omega \epsilon^{ }_2)}
           {m_1^2 - m_2^2 - M^2 + 2 (k p^{ }_2 - \omega \epsilon^{ }_2)}
    \biggr|
  \biggr]
%%% 
\nn 
 & + & 
  n^{ }_{\sigma_2}(\epsilon^{ }_2 - \mu^{ }_2) \, 
  \biggl[
     \frac{\omega k p^{ }_2}{2}
%%% 
\nn 
  & + &   
    \frac{\omega(m_1^2 - m_2^2 - M^2) + 2 \epsilon^{ }_2 M^2}{8} 
    \ln \biggl| 
      \frac{m_1^2 - m_2^2 - M^2 - 2 (k p^{ }_2 - \omega \epsilon^{ }_2)}
           {m_1^2 - m_2^2 - M^2 + 2 (k p^{ }_2 + \omega \epsilon^{ }_2)}
    \biggr|
  \biggr]
%%%%% 
 \biggr\}^{ }_{p^{ }_2 \equiv \sqrt{\epsilon_2^2 - m_2^2}}
%%%%%%%%%%%%%%
  \;\;, \nonumber 
\ea
where 
$
 m^{ }_i \equiv m^{ }_{a_i}
$, 
$
 \mu^{ }_i \equiv \mu^{ }_{a_i}
$, 
and
$
 p^{ }_i \equiv \sqrt{\epsilon_i^2 - m_i^2}
$.

%%%%%%%%%%%%%%%%%%%%%%%%%%% SUBSECTION %%%%%%%%%%%%%%%%%%%%%%%%%%%%%%%%%%%
%
\subsection{Full indirect contribution}
\la{ss:indir_full}

Consider finally the indirect contribution, 
given by \eq\nr{1to2_indir_strctr}. Here the self-energy 
$\bsl{\Sigma}$ includes the 
$2\leftrightarrow 2$ and $1\leftrightarrow 3$ contributions
from \se\ref{se:sterile_2to2},  
the $1+n\leftrightarrow 2+n$ contributions
from \se\ref{se:sterile_1to2}, 
as well as the mass corrections from the real part 
of \eq\nr{1to2_indir_final}. 

With the notation of \eq\nr{master1}, 
the $1\leftrightarrow 2$ contribution to $\bsl{\Sigma}$ is given 
by \eq\nr{1to2_indir_final}, {\it viz.}\
\be
 \aR \bsl{\Sigma} \aL \; \supset \; 
 \aR \, \biggl\{ \, 
 (g_1^2 + g_2^2) \, 
         J^{ -\frac{1}{2} 0 0}_{1\, 1\, 0}
        ( \ana,\aZ, 0 )
 + 
 2 g_2^2 \, 
         J^{ -\frac{1}{2} 0 0}_{1\, 1\, 0}
        ( \aea,\aW, 0 )
 \, \biggr\} \, \aL 
 \;. 
\ee 
For the real part
we only need the contribution proportional to $\msl{\U}$, 
as in \eq\nr{V_def}.
In addition, as alluded to in \se\ref{ss:Rxi},  
at this point it is convenient to include 
the chemical potential representing
a $Z$-tadpole~(ref.~\cite{degenerate}, eq.~(A.7)), 
\be
 \aR \re \bsl{\Sigma} \aL \; \to \; 
 \aR \, \biggl\{ \, 
 (g_1^2 + g_2^2) \, 
         \V^{ -\frac{1}{2} 0 0}_{1\, 1\, 0}
        ( \ana,\aZ,0 )
 + 
 2 g_2^2 \, 
         \V^{ -\frac{1}{2} 0 0}_{1\, 1\, 0}
        ( \aea,\aW,0 )
 - \frac{\muZ}{2}
 \, \biggr\} 
 \, \msl{\U}
 \, 
 \aL 
 \;. \la{muZ}
\ee

We now return to \eq\nr{1to2_indir_strctr}. 
As the self-energy $\bsl{\Sigma}$ includes both a real and
an imaginary part, 
the real and imaginary parts of the inverse read
\ba
 && \hspace*{-1.3cm} 
 \re \Bigl\{ \aR \Bigl[ 
  - \bsl{\K} - \re\bsl{\Sigma} - i \im\bsl{\Sigma}
 \Bigr] \aL  \Bigr\}^{-1}_{ }
 \nn 
 & = & 
 \aL \, 
  \frac{  
 - (\bsl{\K} + \re\bsl{\Sigma})\, 
  \bigl[(\K + \re \Sigma)^2 - (\im\Sigma)^2\bigr]
 - 2 
  \im\bsl{\Sigma}\, 
  ( \K + \re\Sigma ) \cdot\im\Sigma  
  }
 {\bigl[( \K + \re \Sigma)^2 - (\im\Sigma)^2 \bigr]^2
 + \bigl[2 ( \K + \re\Sigma ) \cdot\im\Sigma \bigr]^2}
 \, \aR 
 \;, \la{reProp} \\ 
%%%%%%%%%%%%%%%%%%%%%%%%%%%%%%%%%%%%%
 && \hspace*{-1.3cm} 
 \im \Bigl\{ \aR \Bigl[
  - \bsl{\K} - \re\bsl{\Sigma} - i \im\bsl{\Sigma}
 \Bigr] \aL \Bigr\}^{-1}_{ }
 \nn 
 & = & 
 \aL \, 
  \frac{
  2 (\bsl{\K} + \re\bsl{\Sigma})\, 
  ( \K + \re\Sigma ) \cdot\im\Sigma  
 - 
  \im\bsl{\Sigma}\, 
  \bigl[(\K + \re \Sigma)^2 - (\im\Sigma)^2\bigr] 
  }
 {\bigl[( \K + \re \Sigma)^2 - (\im\Sigma)^2 \bigr]^2
 + \bigl[2 ( \K + \re\Sigma ) \cdot\im\Sigma \bigr]^2}
 \, \aR 
 \;. \la{imProp}
\ea
Before the inversion, we should disentangle the imaginary
part of the self-energy just like the real 
part in \eq\nr{V_def}~\cite{weldon}, {\it viz.}\ 
\be
 \aR \re\bsl{\Sigma} \aL \; = \;
 \aR \bigl(  a \, \bsl{\K} + b \, \msl{\U} \bigr) \aL 
 \;, \quad
 \aR \im\bsl{\Sigma} \aL \; = \; \frac{1}{2}
 \aR \Bigl( 
   \Gamma^{ }_{\!\K} \, \bsl{\K}
 +  
   \Gamma^{ }_{\!\U} \, \msl{\U} 
 \Bigr) \aL 
 \;. \la{def_Sigma}
\ee
Matrix elements taken according to \eqs\nr{Theta_E} and \nr{hel_M}
then reduce to the corresponding matrix elements of 
$\bsl{\K}$ and $\msl{\U}$, 
which are straightforward to evaluate.

%%%%%%%%%%%%%%%%%%%%%%%%%%% SECTION %%%%%%%%%%%%%%%%%%%%%%%%%%%%%%%%%%%%%%
\section{Conclusions}
\la{se:x_concl}

The purpose of the present review has been to collect together 
the theoretical background 
for computing the thermal rate coefficients and mass corrections that
parametrize rate equations for sterile neutrinos in 
the early universe
at $\rmO(h_\nu^2)$ in neutrino Yukawa couplings, while 
keeping the mass scale $M$ and temperature $T$ general. 
This problem originates from a minimal extension of the Standard Model
according to \eq\nr{L}, and
may play a physical role in certain dark matter and 
leptogenesis scenarios. For dark matter, temperatures
much below the electroweak crossover (deep in the Higgs phase) 
are relevant as well. 

In particular, 
we have verified the gauge independence of the
appropriate retarded correlator
(cf.\ \eqs\nr{def_Q} and \nr{def_U}), 
both in the symmetric
and in the Higgs phase of the electroweak theory, 
and after incorporating all 
$1\leftrightarrow 2$, 
$2\leftrightarrow 2$ and 
$1\leftrightarrow 3$ processes. 
We have shown how in the Higgs phase
a subset of the effects can subsequently  
be resummed into an ``indirect contribution'', even though 
this is far from obvious at first sight, given that on the amplitude
level there are interference terms 
between direct and indirect sets
(cf.\ \fig\ref{fig:1to3}). 
The indirect contribution 
dominates at low temperatures, 
and amounts to processes where active 
neutrinos interact with the plasma and are then converted into 
sterile ones. The computations have included generic momenta and 
chemical potentials, and permit for arbitrary helicity projections. 

The missing step of the program would be a numerical evaluation
of all rate coefficients. 
Even though tools for this 
have become available recently~\cite{phasespace,lpm}, 
which have been demonstrated to be 
adaptable to a broad range of problems~\cite{gravity}, 
their practical implementation 
to the current theory
represents a challenge of its own. 
One reason is that the equations 
incorporating LPM-resummation~\cite{lpm} 
are inhomogeneous differential equations and can be time-consuming
in practice. 
Another is 
that the number of objects to be considered ``proliferates'': 
we need to determine both the real and imaginary part of the retarded 
correlator; 
for both helicity states;
for both the direct and indirect contributions. 
Thus there are many observables to determine;
all of them as functions of $M,T,\mu^{ }_i,k$, 
where $\mu^{ }_i$ stands for the set of 
chemical potentials and $k$ is a momentum;
in broad numerical ranges
(say, 
$M = 10^{1  ... 3}$~GeV;
$T = 10^{-4 ... 4}$~GeV;
$|\mu^{ }_i| = 10^{-10 ... -5} T$; 
$k = 10^{-2 ... 2} T$). 
The numerical integrations become demanding
if some particle mass
is much lighter or heavier than the temperature, 
and therefore cannot be performed ``on the fly''. 
Rather, the results are to be tabulated on a dense grid 
in the aforementioned parameter space, suitable for a subsequent
interpolation.
We hope to attack this challenge in the future, noting that 
partial results in a similar spirit have already been 
collected on the web pages 
associated with refs.~\cite{interpolation,broken,cpnumerics}. 

%%%%%%%%%%%%%%%%%%%%%%%%%%% SECTION %%%%%%%%%%%%%%%%%%%%%%%%%%%%%%%%%%
%
\section*{Acknowledgements}

I am grateful to 
Takehiko Asaka, 
Dietrich B\"odeker, 
Jacopo Ghiglieri, 
Ioan Ghisoiu,
Pilar Hern\'andez,  
Greg Jackson, 
York Schr\"oder and
Misha Shaposhnikov for collaborations and discussions over the years
that paved the way for this exposition. 
My work was supported by the Swiss National Science Foundation
(SNSF) under grant 200020B-188712.

%%%%%%%%%%%%%%%%%%%%%%% APPENDIX %%%%%%%%%%%%%%%%%%%%%%%%%%%%%%%%%%%
%
\appendix
\renewcommand{\thesection}{\Alph{section}} % {Appendix~\Alph{section}}
\renewcommand{\thesubsection}{\Alph{section}.\arabic{subsection}}
\renewcommand{\theequation}{\Alph{section}.\arabic{equation}}
%
%%%%%%%%%%%%%%%%%%%%%%%%%%%%%%%%%%%%%%%%%%%%%%%%%%%%%%%%%%%%%%%%%%%%%%%%%%%

%%%%%%%%%%%%%%%%%%%%%%%%%%% SECTION %%%%%%%%%%%%%%%%%%%%%%%%%%%%%%%%%%%%%%
%
\section{$1\to3$ matrix elements squared}
\la{app:1to3}

%%%%%%%%%%%%%%%%%%%%%%%%%%% SUBSECTION %%%%%%%%%%%%%%%%%%%%%%%%%%%%%%%%%%%
%
\subsection{Notation}

In this appendix we specify the matrix elements squared that correspond
to $1\to3$ decays of right-handed neutrinos of mass $M$ into Standard
Model particles. 
As discussed in \ses\ref{ss:crossing} and \ref{ss:virtuals}, 
the same matrix elements also fix, 
via crossing symmetries, the thermal interaction rates originating from 
$2 \leftrightarrow 2$ and $3\to 1$ processes as well as 
those from virtual corrections. 

In order to display the results, we make use of 
Mandelstam variables, {\em viz.}\ % from \eq\nr{sij} 
\be
 \s{12}^{ } \;\equiv\; (\P^{ }_1 + \P^{ }_2)^2 \;,\quad 
 \s{13}^{ } \;\equiv\; (\P^{ }_1 + \P^{ }_3)^2 \;,\quad 
 \s{23}^{ } \;\equiv\; (\P^{ }_2 + \P^{ }_3)^2 \;,
\ee
which are related through
\be
 \s{12}^{ } + \s{13}^{ } + \s{23}^{ } = \MM + m_1^2 + m_2^2 + m_3^2 \;.
 \la{relation}
\ee
Here $m^{ }_i$ denote the masses of the decay products. 
The 
U$^{ }_\rmii{Y}(1)$ and SU$^{ }_\rmii{L}(2)$
gauge couplings $g^{ }_1, g^{ }_2$ are often represented through
\be
 \tilde g^2 \; \equiv \; g_1^2 + g_2^2 
 \;, \quad 
 \sW^2 \; \equiv \; \frac{g_1^2}{\tilde g^2}
 \;, \quad 
 \cW^2 \; \equiv \; \frac{g_2^2}{\tilde g^2}
 \;, \quad 
 \cWW^{ } \; \equiv \; \cW^2 - \sW^2 
 \;. 
\ee

In the symmetric phase, the only non-vanishing mass 
(apart from $M$) is that associated with scalar particles, 
and is denoted by $\mS^{ }$. In the Higgs phase, most particles
carry masses, however those of the charged leptons 
and active neutrinos have been omitted
for simplicity (the same concerns lepton Yukawa 
couplings). By $\aQ$ we denote a massless neutral gauge boson, 
specifically photon in the Higgs phase. 
In the case of charged gauge bosons, 
the symbol $W$ denotes $W_{ }^+$, $\AW$ denotes $W_{ }^-$.
Antiparticles are sometimes 
denoted with a minus sign, e.g.\ $-W \equiv \AW$, 
as they come with opposite chemical potentials. 

For the representation of the matrix elements squared, 
we make use of \eq\nr{Proj}, {\it viz.}
\be
 \Proj \; \equiv \; \frac{\E}{M^2} - \frac{2\E\cdot\K\, \K}{M^4}
 \;. \la{Proj2}
\ee
Then ``direct'' contributions, 
which do not involve the insertion of a Higgs vev, 
are expressed in terms of $\E$, whereas
``indirect contributions'', proportional to $1/M^4$ or $1/M^2$, 
are expressed in terms of $\Proj$. This can be achieved
by employing the relation 
\be
 \E\cdot \P^{ }_i = - \MM \, \Proj\cdot(\P^{ }_j + \P^{ }_k)
                  + (\s{jk}^{ } - m_i^2)\, \Proj\cdot\K 
 \;, \quad
 j,k \neq i \la{EtoI}
\ee
in terms multiplied by $1/\MM$. It turns out that powers of $1/\MM$
come with coefficients $c^{ }_n$ such that 
$
 \sum c^{ }_n (\s{jk}^{ } - m_i^2) = \rmO(\MM) 
$, 
so that \eq\nr{EtoI} eliminates the appearance of $1/\MM$.
Going in the opposite direction, $\Proj$'s not multiplied by 
a Higgs vev can be eliminated by
\be
 M^4\, \Proj\cdot \P^{ }_i = - \MM\,\E\cdot(\P^{ }_j + \P^{ }_k)
                  + (\s{jk}^{ } - m_i^2)\, \E\cdot\K 
 \;, \quad
 j,k \neq i
 \;.  \la{ItoE}
\ee

Unfortunately, the splitup into $\E$ and $\Proj$ is not unique.
This should not be surprising, given that 
the direct and indirect sets are not gauge independent by themselves. 
In addition, powers of~$\MM$ cannot be identified uniquely, 
as kinematic variables are related through \eq\nr{relation}. 
We have attempted to employ this freedom 
in order to express the results 
in a compact fashion, however the procedure has ambiguities. 
Furthermore, the appearance of kinematic variables in the numerator
leads to several alternative representations.
In the end, to verify the equivalence of two representations, 
one needs to choose some lexicographic ordering and an algorithm putting 
the expression in this form, which leads generically to 
lengthier expressions, but unique ones in the chosen basis.  

We express the result as a contribution to $\Theta^{ }_{\E}$
from \eq\nr{Theta_E}. In terms of matrix elements, 
the results correspond to 
$
 \sum |\mathcal{M}|^2
$, 
with all final-state spins summed over.
For the right-handed states, \eq\nr{Theta_E}
implies that we need to adopt the effective replacement
\be
 \sum_\tau u^{ }_{\vec{k}\tau} \bar{u}^{ }_{\vec{k}\tau} 
 \longrightarrow
 \msl{\E}
 \;, \la{replace} 
\ee
with $\E$ fixed according to \eq\nr{hel_M}.
For thermal phase space averaging we make use of 
the notation in \eqs\nr{x_scat1to3a}--\nr{x_scat1to3c}.

The expressions for the matrix elements squared enjoy certain symmetries 
that provide for stringent crosschecks of their correctness. This concerns
particular the interference terms, which necessarily appear in the 
structure of \eq(2.34) of ref.~\cite{phasespace}, {\it viz.} 
\ba
 \Theta^{ }_{\E} & \supset & 
 \scat{1\to3}(a,b,c) \, 
 \frac{\widetilde{\Theta}^{ }_{ }(\P^{ }_a,\P^{ }_b,\P^{ }_c)}
 {[\,(\P^{ }_a + \P^{ }_b)^2 - m_d^2\,]\,
  [\,(\P^{ }_c + \P^{ }_b)^2 - m_e^2\,]}
%%%%%%%%%%%%%
 \nn 
 & + & 
 \scat{1\to3}(d,-b,e) \, 
 \frac{\widetilde{\Theta}^{ }_{ }
  (\P^{ }_d+\P^{ }_b,-\P^{ }_b,\P^{ }_e+\P^{ }_b)}
 {[\,(\P^{ }_d + \P^{ }_b)^2 - m_a^2\,]\,
  [\,(\P^{ }_e + \P^{ }_b)^2 - m_c^2\,]}
 \;. \la{structure_master}
\ea
By making use of \eq\nr{EtoI}, it can be verified that the 
interference terms in \eqs\nr{leptonic_symmetric} and 
\nr{leptonic_broken} indeed satisfy this property. 

Machine-readable versions of 
\eqs\nr{hadronic_symmetric}--\nr{leptonic_broken}
are attached to this paper as ancillary files. 

%%%%%%%%%%%%%%%%%%%%%%%%%%% SUBSECTION %%%%%%%%%%%%%%%%%%%%%%%%%%%%%%%%%%%
%
\subsection{Hadronic effects, symmetric phase}

In the symmetric phase, decays into final states containing hadrons amount to
{\small
\ba
 \Theta^{ }_{\E}
 & \supset & 
                \scat{1\to3}(\ala, \AtopL,\atopR)12\,\htop^2
                 \,\biggl[ 
% \nn & & 
                  \frac{ \s{23}^{ }
                          \,\E\cdot\P^{ }_1
                         }{ ({ \Sq{\s{23}^{ }}{\mS} })^2 } 
                  \biggr]
%**************************
 \nn & + & 
                \scat{1\to3}(\ala, \AbotR,\atopL)12\,\hbot^2
                 \,\biggl[ 
% \nn & & 
                  \frac{ \s{23}^{ }
                          \,\E\cdot\P^{ }_1
                          }{ ({ \Sq{\s{23}^{ }}{\mS} })^2 }
                  \biggr]
 \;. \la{hadronic_symmetric}
\ea}\normalsize%
Here $h^{ }_t$ and $h^{ }_b$ are the top and bottom Yukawa couplings, 
the others having been omitted. 

%%%%%%%%%%%%%%%%%%%%%%%%%%% SUBSECTION %%%%%%%%%%%%%%%%%%%%%%%%%%%%%%%%%%%
%
\subsection{Hadronic effects, Higgs phase}

When the Higgs mechanism is active, both direct and indirect 
contributions are present, and many different masses play a role. 
In this section we display the processes involving hadronic final states. 
It can be checked that for $m^{ }_i\to 0$ the Higgs phase results
go over into \eq\nr{hadronic_symmetric}, when the latter
is evaluated with $\mS\to 0$. 
For simplicity we keep only the top and bottom quark masses
finite ($\mtop,\mbot$), whereas the other quarks are treated as
massless, with furthermore the labels set to those of the 
up and down quark, {\it viz.} $c\to u$, $s\to d$.
Then the results read
{\small 
\ba
 \Theta^{ }_{\E} 
 &  \supset & 
          \scat{1\to3}(\ana, \Atop,\atop)\,\gt^2\,\biggl\{ 
 \nn & & 
                   + \frac{ 
                      \mZ^2\,
                      [
                        8 \sW^2(3-4\sW^2) \mtop^2 \, \Proj\cdot \P^{ }_1
                    - (3-4\sW^2)^2 (\s{13}^{ } - \mtop^2)\, \Proj\cdot \P^{ }_2
                     - 16 \sW^4 (\s{12}^{ } - \mtop^2)\, \Proj\cdot \P^{ }_3
                      ]
                    }{ 
                       3 ({ \Sq{\s{23}^{ }}{\mZ} })^2 } 
 \nn & & 
                   + \frac{ 
                      (3-8\sW^2) \mtop^2
                      [  (\s{12}^{ } - \mtop^2) \, \Proj\cdot \P^{ }_2
                       - (\s{13}^{ } - \mtop^2) \, \Proj\cdot \P^{ }_3 
                      ]
                    }{  ({ \Sq{\s{23}^{ }}{\mZ} })
                        ({ \Sq{\s{23}^{ }}{\mh} }) }
 \nn & & 
                   + \frac{ 3 \mtop^2 \,\E\cdot\P^{ }_1 }{ 2 \mZ^2 }
                    \biggl[ 
                    \frac{\s{23}^{ } - 2 \mZ^2}
                         { ({ \Sq{\s{23}^{ }}{\mZ} })^2 }
                   + 
                    \frac{ 
                     \s{23}^{ }  - 4 \mtop^2
                    }{
                         ({ \Sq{\s{23}^{ }}{\mh} })^2 }
                    \biggr]
                   \biggr\} 
%******************************************************
 \nn[3mm] & + & 
                 \scat{1\to3}(\ana, \Aup,\aup)\, 
                 \, 2\gt^2\mZ^2\,\biggl\{  
% \nn & & +
                  - \frac{ 
                        (3-4\sW^2)^2  \s{13}^{ }  
                       \,\Proj\cdot\P^{ }_2
                  + 
                       16 \sW^4  \s{12}^{ } 
                       \, \Proj\cdot \P^{ }_3
                        } 
                   { 3 ({ \Sq{\s{23}^{ }}{\mZ} })^2 }
                   \biggr\} 
%******************************************************
 \nn & + & 
                 \scat{1\to3}(\ana, \Abot,\abot)\,\gt^2\,\biggl\{ 
 \nn & & 
                   + \frac{  
                       \mZ^2 \, 
                              [
                               4 \sW^2(3-2\sW^2) \mbot^2 \,  \Proj\cdot\P^{ }_1
                   - (3-2\sW^2)^2(\s{13}^{ } - \mbot^2) \, \Proj\cdot\P^{ }_2
                           - 4\sW^4(\s{12}^{ } - \mbot^2) \, \Proj\cdot\P^{ }_3
                              ]
                    }{
                        3 ({ \Sq{\s{23}^{ }}{\mZ} })^2 }
 \nn & & 
                   - \frac{
                     (3-4 \sW^2)
                     \mbot^2
                     [ (\s{12}^{ } - \mbot^2) \, 
                            \Proj\cdot \P^{ }_2
                      - 
                       (\s{13}^{ } - \mbot^2 ) \, 
                            \Proj\cdot\P^{ }_3
                     ]
                    }{ ({ \Sq{\s{23}^{ }}{\mZ} })
                       ({ \Sq{\s{23}^{ }}{\mh} })  }
 \nn & & 
                   + \frac{ 3 \mbot^2 \,\E\cdot\P^{ }_1 }
                          { 2 \mZ^2 } 
                     \biggl[ \frac{\s{23}^{ } - 2 \mZ^2}
                                  { ({ \Sq{\s{23}^{ }}{\mZ} })^2 }
                      + \frac{
                          \s{23}^{ } - 4 \mbot^2 
                     }{ 
                        ({ \Sq{\s{23}^{ }}{\mh} })^2 }
                     \biggr]
                   \biggr\} 
%******************************************************
 \nn[3mm] & + & 
                  \scat{1\to3}(\ana, \Ado,\ado)
                  \, 2 \gt^2 \mZ^2 \,\biggl\{ 
% \nn & & + 
                   - \frac{ 
                         (3 - 2\sW^2)^2  
                         \, \s{13}^{ } \, \Proj\cdot\P^{ }_2
                        + 
                          4\sW^4 \, \s{12}^{ } \, \Proj\cdot \P^{ }_3 
                     }{ 
                        3 ({ \Sq{\s{23}^{ }}{\mZ} })^2 }
                   \biggr\} 
%******************************************************
 \nn & + & 
                  \scat{1\to3}(\aea, \Abot,\atop)
                  \,3g_2^2\, |V^{ }_{\!tb}|^2 \, \biggl\{ 
 \nn & & 
                   - \frac{ 
                     2\mW^2 
                    [
                      (\s{13}^{ } - \mtop^2)
                      \,\Proj\cdot\P^{ }_2
                     + 
                      (\s{12}^{ } - \mbot^2)
                      \,\Proj\cdot\P^{ }_3
                    ]
                         }{ 
                        ({ \Sq{\s{23}^{ }}{\mW} })^2 }
 \nn & & 
                   + \frac{ 
                     2 (\mtop^2 - \mbot^2) \, 
                      [
                         (\mtop^2 - \mbot^2) \,\Proj\cdot\P^{ }_1
                       + (\s{12}^{ } - \mbot^2 )  \,\Proj\cdot\P^{ }_2
                       - (\s{13}^{ } - \mtop^2 )  \,\Proj\cdot\P^{ }_3
                      ]
                         }{ 
                        ({ \Sq{\s{23}^{ }}{\mW} })^2 }
 \nn & & 
                   + \frac{ 
                      [
                         (\mtop^2 + \mbot^2) 
                         ( \s{23}^{ } - 2 \mW^2 ) 
                       - (\mtop^2 - \mbot^2)^2 ] 
                         \, \E \cdot \P^{ }_1
                      ]
                    }{
                        \mW^2 ({ \Sq{\s{23}^{ }}{\mW} })^2 }
                   \biggr\}
%******************************************************
 \nn[3mm] & + & 
                  \scat{1\to3}(\aea, \Ado,\aup)
                   \, 3g_2^2 \mW^2
                   \, \bigl(  |V^{ }_{\!ud}|^2 + |V^{ }_{\!us}|^2
                            + |V^{ }_{\!cd}|^2 + |V^{ }_{\!cs}|^2   
                      \bigr)
                   \, \biggl\{ 
 \nn & &  
                    - \frac{ 
                      2  
                      [ 
                      \s{13}^{ }
                      \, \Proj \cdot\P^{ }_2
                      + 
                      \s{12}^{ }
                      \, \Proj \cdot\P^{ }_3
                      ]
                    }{ ({ \Sq{\s{23}^{ }}{\mW} })^2 }
                   \biggr\} 
 \;. \la{hadronic_broken}
\ea}\normalsize\renewcommand{\baselinestretch}{1.15}%
The last two terms, originating from charged currents, have been 
simplified slightly, suppressing small off-diagonals from 
the CKM matrix elements $ |V^{ }_{\!td}|^2 $, $ |V^{ }_{\!ts}|^2 $, 
$ |V^{ }_{\!ub}|^2 $, $ |V^{ }_{\!cb}|^2 $ as well as parity-odd
contributions involving the Levi-Civita tensor. 

%%%%%%%%%%%%%%%%%%%%%%%%%%% SUBSECTION %%%%%%%%%%%%%%%%%%%%%%%%%%%%%%%%%%%
%
\subsection{Leptonic effects, symmetric phase}

In the symmetric phase, the processes 
only involving leptons in the final state 
amount to 
{\small
\ba
 \Theta^{ }_{\E}
 & \supset &  \scat{1\to3}(\ell^{ }_a,\aQ,\aS)\, 2 ( g_1^2 + 3 g_2^2 )
                \biggl\{ 
 \nn & & 
                 -    \frac{\E\cdot\P^{ }_1}{ ({ \s{12}^{ } })}
                 +     \frac{
                       (\MM - \mS^2)\,
                        \E\cdot( 2 \P^{ }_1 +  \P^{ }_2 )  
                       }{ 
                        ({ \s{12}^{ } }) ({ \Sq{\s{23}^{ }}{\mS} }) }
%  \nn & &  
                 -      \frac{ 
                          \E\cdot(\P^{ }_1 + \K )    
                             }{ ({ \Sq{\s{23}^{ }}{\mS} }) }
                 -      \frac{ 
                         2 \mS^2 \, \E\cdot\P^{ }_1   
                             }{ ({ \Sq{\s{23}^{ }}{\mS} })^2 }
                       \biggr\} 
 \;. \la{leptonic_symmetric}
\ea}\normalsize
Here $\aS$ denotes a scalar field and $\aQ$ a massless gauge boson. 

%%%%%%%%%%%%%%%%%%%%%%%%%%% SUBSECTION %%%%%%%%%%%%%%%%%%%%%%%%%%%%%%%%%%%
%
\subsection{Leptonic effects, Higgs phase}

In the Higgs phase, \eq\nr{leptonic_symmetric} splits into a large
number of contributions, and new indirect processes proportional to $\Proj$
are generated. However, 
it can be checked that for $m^{ }_i\to 0$ 
only terms of the type $\E\cdot \P^{ }_i$ survive and 
the Higgs phase results
go over into \eq\nr{leptonic_symmetric}, evaluated with $\mS\to 0$
(the sterile neutrino mass $M$ is non-zero in both cases).
The full list of Higgs phase results, which is independent
of the gauge fixing parameter,  reads
{\small 
\ba
 \Theta^{ }_{\E}
 & \supset &
 \scat{1\to3}(\ana, \AW,\aW)\,g_2^2\,\biggl\{
 \nn & & 
          - \frac{ 
                 (\mZ^2-4\mW^2)
                 \, [ 8\mW^2 (\s{23}^{ } + \mZ^2)\,\Proj\cdot\P^{ }_1
                 -(2\s{23}^{ } - \mZ^2 + 8\mW^2)\,\E\cdot\P^{ }_1 ]
                 -12 \mW^4\,\E\cdot\P^{ }_1
                 }{ 4 \mW^2 ({ \Sq{\s{23}^{ }}{\mZ} })^2 } 
 \nn & & 
          + \frac{ 
                 (\mZ^4-4\mW^2\mZ^2+12\mW^4)
                 \,[\, (\s{13}^{ } + \mW^2)
                         \,\Proj\cdot\P^{ }_1
                     + (2\s{13}^{ } + \mZ^2-2\mW^2)
                         \,\Proj\cdot\P^{ }_2
                 \,]\,
                 }{ 2 \mW^2 ({ \Sq{\s{23}^{ }}{\mZ} })^2 } 
 \nn & & 
          + \frac{ 
                  [\, (\mZ^2-4\mW^2)\s{23}^{ }
                 +12\mW^4\,]\,\E\cdot\P^{ }_3
                 }{ 2 \mW^2 ({ \Sq{\s{23}^{ }}{\mZ} })^2 } 
%%%%%%%%%%%%%%%%%%%%%%%%%%%%
% \nn & & 
          + \frac{ 
                  (\mh^4-4\mW^2\s{23}^{ } + 12\mW^4)\,\E\cdot\P^{ }_1
                 }{ 4\mW^2 ({ \Sq{\s{23}^{ }}{\mh} })^2 } 
%%%%%%%%%%%%%%%%%%%%%%%%%%%%
 \nn & & 
          - \frac{ 
                  (\mZ^2-2\mW^2) \s{13}^{ }  
                          [ (\mh^2 -\mZ^2)
                              \,\Proj\cdot \K  
                             +  \mh^2 
                             \,\Proj\cdot (\P^{ }_2 + \P^{ }_3)\,]
           }{ 2\mW^2 ({ \Sq{\s{23}^{ }}{\mZ} }) ({ \Sq{\s{23}^{ }}{\mh} }) } 
 \nn & & 
          + \frac{ 
                  (\mZ^2-6\mW^2) \mW^2 
                         [ (\mh^2-\mZ^2)
                                   \,\Proj\cdot(\P^{ }_1 + 2\P^{ }_3)
                          -2\s{13}^{ }\,\Proj\cdot(\P^{ }_2 + \P^{ }_3) 
                         \,]
            }{ 2\mW^2 ({ \Sq{\s{23}^{ }}{\mZ} }) ({ \Sq{\s{23}^{ }}{\mh} }) } 
 \nn & & 
          - \frac{ (\mZ^4 - 12 \mW^4) 
                         [ (\mZ^2 - \mW^2 - \MM) \,\Proj\cdot\P^{ }_2
                           - \mW^2 \, \Proj\cdot \P^{ }_3 
                         \,] 
           }{ 2 \mW^2 ({ \Sq{\s{23}^{ }}{\mZ} }) ({ \Sq{\s{23}^{ }}{\mh} }) } 
%%%%%%%%%%%%%%%%%%%%%%%%%%%%
 \nn & & 
          + \frac{
                   4 \mW^2   
                    [ 
                       \s{12}^{ }\,\Proj\cdot\P^{ }_1
                      + (\s{12}^{ }+\mW^2)\,\Proj\cdot(\P^{ }_1 + \P^{ }_2)
                    ]
                  - 2 [\, \s{12}^{ }\,\E\cdot\P^{ }_1
                   + \mW^2   
                  \, \E \cdot(\P^{ }_1 + \P^{ }_2)
                       \,]
                 }{ ({ \s{12}^{2} }) } 
%%%%%%%%%%%%%%%%%%%%%%%%%%%%
 \nn & & 
          + \frac{ 
                  2(5\mZ^2+7\mW^2)\mW^2\,\Proj\cdot\P^{ }_1
                 + (2 \MM - \mZ^2 - \s{12}^{ })\,\E\cdot\P^{ }_1
                 }{ ({ \s{12}^{ } }) ({ \Sq{\s{23}^{ }}{\mZ} }) } 
 \nn & & 
          + \frac{ 
                 [
                   (\mZ^2-2\mW^2)^2 \s{12}^{ } 
                   + 4(2\mZ^2+\mW^2)\mW^4 
                   ] \,\Proj\cdot\P^{ }_2
                 +(\mZ^2+ 3\mW^2 +\MM - \s{12}^{ }) \mW^2 \,\E\cdot\P^{ }_2
               }{ \mW^2 ({ \s{12}^{ } }) ({ \Sq{\s{23}^{ }}{\mZ} }) } 
 \nn & & 
          + \frac{ 
                  2[
                   \mZ^2\s{23}^{ } 
                  +(\mZ^2-2\mW^2)\mW^2 
                  ] \,\Proj\cdot\P^{ }_3 
                 +9\mW^2\,\E\cdot\P^{ }_3
                 }{ ({ \s{12}^{ } }) ({ \Sq{\s{23}^{ }}{\mZ} }) } 
%%%%%%%%%%%%%%%%%%%%%%%%%%%%
 \nn & & 
          + \frac{ 
                   4\mW^6\,\Proj\cdot \P^{ }_1  
              - [\mZ^2\s{12}^{ } + 2\mW^2( \mh^2 - 2\MM) ]\,\E\cdot\P^{ }_1
             }{ 2 \mW^2 ({ \s{12}^{ } }) ({ \Sq{\s{23}^{ }}{\mh} }) } 
 \nn & & 
          - \frac{ 
                 [ 
                    (\mZ^2-2\mW^2)(\mh^2 + \mZ^2) \s{12}^{ }
                   - 8 \mW^6
                 ] \,\Proj\cdot\P^{ }_2
                 +2 (\s{12}^{ } + \mh^2+\mW^2-\MM )\mW^2\,\E\cdot\P^{ }_2
             }{ 2 \mW^2 ({ \s{12}^{ } }) ({ \Sq{\s{23}^{ }}{\mh} }) } 
 \nn & & 
          - \frac{ 
                  4 (\mh^2 - 2\mW^2)\mW^4\,\Proj\cdot\P^{ }_3
                 +(\mZ^2\s{12}^{ } - 2\mW^4)\,\E\cdot\P^{ }_3
             }{ 2 \mW^2 ({ \s{12}^{ } }) ({ \Sq{\s{23}^{ }}{\mh} }) } 
                   \biggr\} 
%\******************************************************
 \nn & + & 
                   \scat{1\to3}(\ana,\aZ,\aZ)\,\gt^2\,\biggl\{ 
 \nn & & 
                   + \frac{ 
                      \mZ^2[
                            (\s{12}^{ } + \mZ^2) \, \Proj\cdot\P^{ }_1
                          + (\mZ^2-\s{12}^{ }) \, \Proj\cdot\P^{ }_2  
                          + (\s{12}^{ }-\mZ^2+\MM) \, \Proj\cdot\P^{ }_3
                           ]
                     }{ 
                        2 ({ \s{12}^{2} }) }
%%%%%%%%%%%%%%%%%%%%%%%%%
 \nn & & 
                   + \frac{ 
                      (\mh^4 - 4\mZ^2\mh^2+12\mZ^4)  \, \E \cdot\P^{ }_1 
                    }{
                       8 \mZ^2 ({ \Sq{\s{23}^{ }}{\mh} })^2 }
%%%%%%%%%%%%%%%%%%%%%%%
% \nn & & 
                   - \frac{ 
                      \mZ^2\, (2\mZ^2 - \MM) \, \Proj\cdot \P^{ }_1 
                    }{
                      ({ \s{12}^{ } }) ({ \s{13}^{ } }) }
%%%%%%%%%%%%%%%%%%%%%%%%%
 \nn & & 
                   - \frac{ 
                      \mZ^2\, [\, 
                        (\s{12}^{ } - \mZ^2)   \, \Proj\cdot \P^{ }_2
                      + (\mZ^2 - \s{13}^{ })   \, \Proj\cdot \P^{ }_3
                            \,]
                    }{
                     ({ \s{12}^{ } }) ({ \Sq{\s{23}^{ }}{\mh} }) }
%%%%%%%%%%%%%%%%%%%%%%%%%
 \nn & & 
                   - \frac{ 
                          (\s{12}^{ }-\s{13}^{ }  - \MM) \, \E\cdot\P^{ }_1 
                     + (\s{12}^{ } + \mh^2 - \mZ^2 - \MM) \, \E\cdot\P^{ }_2
                     + (\s{12}^{ } - \mZ^2) \, \E\cdot\P^{ }_3
                    }{
                     2 ({ \s{12}^{ } }) ({ \Sq{\s{23}^{ }}{\mh} }) }
                   \biggr\} 
%******************************************************
 \nn & + & 
                  \scat{1\to3}(\ana,\aZ,\ah)\,\gt^2\,\biggl\{ 
 \nn & & 
                   - \frac{ 
                       \mZ^2
                      [\, (2 \mZ^2 - \mh^2) \, \Proj\cdot\P^{ }_1
                       + (\s{13}^{ } - \mh^2) \, \Proj\cdot\P^{ }_3
                             \,]
                     }{ 
                        ({ \Sq{\s{23}^{ }}{\mZ} })^2 }
 \nn & & 
                   - \frac{ 
                      \mh^2 
                      [\, (\mh^2 - \mZ^2) \, \Proj\cdot\P^{ }_1
                       + (\s{12}^{ } - \mZ^2) \, \Proj\cdot\P^{ }_2
                       + (\mh^2 - \s{13}^{ }) \, \Proj\cdot\P^{ }_3
                             \,]
                     }{ 
                       2 ({ \Sq{\s{23}^{ }}{\mZ} })^2 }
 \nn & & 
                   + \frac{
                       (4\mZ^4 -2\mZ^2\mh^2 + \mh^4) 
                               \, \E \cdot\P^{ }_1
                     }{ 
                       4\mZ^2 ({ \Sq{\s{23}^{ }}{\mZ} })^2 }
% \nn & & 
                   - \frac{ 
                         \mZ^2\, \E\cdot ( \P^{ }_1 + \P^{ }_2 )
                     }{ 
                        2 ({ \s{12}^{2} })   }
 \nn & & 
                   - \frac{  
                       \mZ^2 [\, 
                       (\s{12}^{ } -\s{13}^{ } + \MM )\, \Proj \cdot\P^{ }_1 
                      +(\mh^2 -\s{13}^{ }) \, \Proj\cdot\P^{ }_2
                      +(\s{12}^{ } - \mZ^2) \, \Proj\cdot\P^{ }_3 \,] 
                    }{ 
                      ({ \s{12}^{ } }) ({ \Sq{\s{23}^{ }}{\mZ} }) }
 \nn & & 
                   - \frac{  
                   (\s{12}^{ } - \s{13}^{ } + 2 \mh^2 - \MM)\, \E\cdot\P^{ }_1
                    + (\s{12}^{ } - \MM ) \,\E\cdot\P^{ }_2 
                    + (\s{12}^{ } - \mZ^2) \,\E\cdot\P^{ }_3
                    }{ 
                      2 ({ \s{12}^{ } }) ({ \Sq{\s{23}^{ }}{\mZ} }) }
                   \biggr\} 
%******************************************************
 \nn[3mm] & + & 
                   \scat{1\to3}(\ana,\ah,\ah)\,\lam\,\biggl\{ 
% \nn & & 
                  + \frac{ 
                     9 \mh^2\, \E\cdot\P^{ }_1
                   }{ ({ \Sq{\s{23}^{ }}{\mh} })^2 }
                   \biggr\}
%******************************************************
 \nn & + & 
                   \scat{1\to3}(\aea,\aW,\aZ)\,\gt^2\mW^2\,\biggl\{ 
 \nn & & 
                  + \frac{
                     [2(8-15\sW^2+4\sW^4)(\s{23}^{ } + \mW^2) 
                     +(9 - 8\sW^2) \s{13}^{ }
                     + 2\mW^2 - \mZ^2 
                     ] \, \Proj\cdot\P^{ }_1
                        }{
                           ({ \Sq{\s{23}^{ }}{\mW} })^2 }
 \nn & & 
                  + \frac{
                             [\,  
                       2(9 - 8\sW^2 ) 
                       (\MM - \s{12}^{ }) 
                        - 9 \s{23}^{ } +\mZ^2 + 17 \mW^2   
                              \, ] \, \Proj\cdot\P^{ }_2
                        }{
                           ({ \Sq{\s{23}^{ }}{\mW} })^2 }
 \nn & & 
                  + \frac{
                        \sW^2\, [\,  
                       8 \cWW^{ } (\s{23}^{ } - \mW^2)  
                     + (9-8\sW^2)(2\s{13}^{ } - \mZ^2) 
                              \, ] \, \Proj\cdot\P^{ }_3
                     + (9 - 8\sW^2) 
                         \, \E\cdot\P^{ }_3
                        }{
                           ({ \Sq{\s{23}^{ }}{\mW} })^2 }
 \nn & & 
                  - \frac{( 15 \cWW^2 - 16\sW^6 ) \, \mZ^2 
                         \, \E\cdot\P^{ }_1
                      + 2 ( \cW^2 + \cWW^2 )(\s{23}^{ } - \mW^2)\, 
                         \E\cdot( \K + \P^{ }_1 ) 
                        }{
                           2 \mW^2 ({ \Sq{\s{23}^{ }}{\mW} })^2 }
%%%%%%%%%%%%%%%%%
 \nn & & 
                  + \frac{
                     2 \mZ^2 \, [
                         \s{12}^{ } \, \Proj\cdot\P^{ }_1
                    + (\s{12}^{ }+\mW^2) \, \Proj\cdot(\P^{ }_1 + \P^{ }_2 )
                       ]
                       - \s{12}^{ } \, \E\cdot\P^{ }_1  
                       - \mW^2 \, \E\cdot(\P^{ }_1 + \P^{ }_2 )
                        }{
                           \mZ^2 ({ \s{12}^{2} }) }
%%%%%%%%%%%%%%%%%
 \nn & & 
                  + \frac{
                     \cWW^2 
                     \{ 2 \mW^2 
                      [ \s{13}^{ } \, \Proj\cdot\P^{ }_1
                      + (\s{13}^{ }+\mZ^2) \, \Proj\cdot(\P^{ }_1 + \P^{ }_3) ]
                         - \s{13}^{ } \, \E\cdot\P^{ }_1
                         - \mZ^2 \, \E\cdot( \P^{ }_1 + \P^{ }_3)
                     \} 
                        }{
                           \mW^2 ({ \s{13}^{2} }) }
%%%%%%%%%%%%%%%%%
 \nn & & 
                  + \frac{
                       2 \mW^2 (5\mZ^2 + 7\mW^2) \, \Proj\cdot\P^{ }_1
                      - (\mZ^2 - 2\MM)\,  \E\cdot\P^{ }_1
                    }{
                 \mZ^2 ({ \s{12}^{ } }) ({ \Sq{\s{23}^{ }}{\mW} }) }
 \nn & & 
                  + \frac{
                    2 (\mZ^4 + 2\mW^2\mZ^2 + 3\mW^4 ) 
                       \, \Proj\cdot\P^{ }_2
                      - (2\mZ^2-6\mW^2 - \MM)\,\E\cdot\P^{ }_2
                    }{
                  \mZ^2 ({ \s{12}^{ } }) ({ \Sq{\s{23}^{ }}{\mW} }) }
 \nn & & 
                  + \frac{
                       2 (\mZ^2 - \mW^2) (\s{23}^{ } + \mW^2 + \mZ^2) 
                       \, \Proj\cdot\P^{ }_3
                       + 9 \mW^2\, \E\cdot\P^{ }_3
                    }{
                  \mZ^2 ({ \s{12}^{ } }) ({ \Sq{\s{23}^{ }}{\mW} }) }
%%%%%%%%%%%%%%%%%
 \nn & & 
                  - \frac{
                      \cWW^{ } [
                       4\mW^2 (\mZ^2 - 7 \mW^2 + 4 \sW^2 \mW^2)
                               \, \Proj\cdot\P^{ }_1
                  + \cWW^{ }
                       (2\mW^2 -  
                       \mZ^2 -2 \MM 
                       )\, \E\cdot\P^{ }_1 
                       ] 
                    }{
            \mW^2 ({ \s{13}^{ } }) ({ \Sq{\s{23}^{ }}{\mW} }) }
 \nn & & 
                  - \frac{
                       \cWW^{ } [ 
                       4\mW^2 \sW^2(\s{23}^{ } + \mW^2 + \mZ^2 )
                               \, \Proj\cdot\P^{ }_2
                        + (\mZ^2 - 10 \mW^2 + 8 \sW^2 \mW^2)
                               \, \E\cdot\P^{ }_2
                                  ] 
                    }{
                \mW^2 ({ \s{13}^{ } }) ({ \Sq{\s{23}^{ }}{\mW} }) }
 \nn & & 
                  - \frac{
                        \cWW^{ } \{  
                        4 \mW^2 
                        ( \mZ^2 - 4\mW^2
                            + 3\sW^2\mW^2)\, \Proj\cdot\P^{ }_3
                       - [ 
                       \cWW^{ } (2\mZ^2+\mW^2+\MM) + \mZ^2 
                       ]
                       \,\E\cdot\P^{ }_3
                       \} 
                    }{
              \mW^2 ({ \s{13}^{ } }) ({ \Sq{\s{23}^{ }}{\mW} }) }
%%%%%%%%%%%%%%%%%
 \nn & & 
                  + \frac{
                       4 \cWW^{ } [ (\mW^2+\mZ^2) 
                       \, \Proj\cdot(\K +  \P^{ }_1)
                       -  \E \cdot\P^{ }_1 ]
                    }{
                      ({ \s{12}^{ } }) ({ \s{13}^{ } }) }
                   \biggr\} 
%******************************************************
 \nn[3mm] & + & 
                   \scat{1\to3}(\aea,\aW,\aQ)\, 4 g_2^2 \sW^2 \, \biggl\{ 
 \nn & & 
                   + \frac{ 
                     2 \mW^2 \,[\,
                         (\s{23}^{ } + \mW^2) \,\Proj\cdot\P^{ }_1
                       +2(\s{12}^{ } - \MM) \,\Proj\cdot\P^{ }_3 \,] 
                     }{
                        ({ \Sq{\s{23}^{ }}{\mW} })^2 }
% \nn & & 
                   - \frac{ 
                     2 \mW^2 \,\E\cdot\P^{ }_1
                     }{
                        ({ \Sq{\s{23}^{ }}{\mW} })^2 }
 \nn & & 
                   + \frac{ 
                     2 \mW^2  \, [ \, 
                         \s{23}^{ } \,\Proj\cdot\P^{ }_1
                        - \mW^2 \,\Proj\cdot\P^{ }_2
                       + (\s{23}^{ } + 2 \mW^2)\,
                      \Proj\cdot\K % (\P^{ }_1 + \P^{ }_2 + \P^{ }_3 )
                              \, ] 
                     }{
                        ({ \s{13}^{ } }) ({ \Sq{\s{23}^{ }}{\mW} }) }
 \nn & & 
                   - \frac{ 
                  (\s{13}^{ } - \s{12}^{ } + 2\mW^2 - \MM )\, \E\cdot\P^{ }_1
                   + (\s{13}^{ } - 4\mW^2 ) \, \E\cdot\P^{ }_2
                   + (\s{13}^{ } - \mW^2 - \MM) \, \E\cdot\P^{ }_3
                     }{
                    ({ \s{13}^{ } }) ({ \Sq{\s{23}^{ }}{\mW} }) }
                   \biggr\} 
%\******************************************************
 \nn & + & 
                  \scat{1\to3}(\aea,\aW,\ah)\,g_2^2 \, \biggl\{  
 \nn & & 
                   - \frac{ 
                      2 \mW^2
                      [\, (2 \mW^2 - \mh^2) \, \Proj\cdot\P^{ }_1
                       + (\s{13}^{ } - \mh^2) \, \Proj\cdot\P^{ }_3
                             \,]
                     }{ 
                        ({ \Sq{\s{23}^{ }}{\mW} })^2 }
 \nn & & 
                   - \frac{ 
                      \mh^2 
                      [\, (\mh^2 - \mW^2) \, \Proj\cdot\P^{ }_1
                       + (\s{12}^{ } - \mW^2) \, \Proj\cdot\P^{ }_2
                       + (\mh^2 - \s{13}^{ }) \, \Proj\cdot\P^{ }_3
                             \,]
                     }{ 
                        ({ \Sq{\s{23}^{ }}{\mW} })^2 }
 \nn & & 
                   + \frac{
                       (4\mW^4 -2\mW^2\mh^2 + \mh^4) 
                               \, \E \cdot\P^{ }_1
                     }{ 
                       2\mW^2 ({ \Sq{\s{23}^{ }}{\mW} })^2 }
% \nn & & 
                   - \frac{ 
                         \mW^2\, \E\cdot ( \P^{ }_1 +  \P^{ }_2 )
                     }{ 
                         ({ \s{12}^{2} })   }
 \nn & & 
                   - \frac{  
                       2\mW^2 [\, 
                       (\s{12}^{ } -\s{13}^{ } + \MM )\, \Proj \cdot\P^{ }_1 
                      +(\mh^2 -\s{13}^{ }) \, \Proj\cdot\P^{ }_2
                      +(\s{12}^{ } - \mW^2) \, \Proj\cdot\P^{ }_3 \,] 
                    }{ 
                      ({ \s{12}^{ } }) ({ \Sq{\s{23}^{ }}{\mW} }) }
 \nn & & 
                   - \frac{  
                 (\s{12}^{ } - \s{13}^{ } + 2 \mh^2 - \MM)\, \E\cdot\P^{ }_1
                    + (\s{12}^{ } - \MM ) \,\E\cdot\P^{ }_2 
                    + (\s{12}^{ } - \mW^2) \,\E\cdot\P^{ }_3
                    }{ 
                   ({ \s{12}^{ } }) ({ \Sq{\s{23}^{ }}{\mW} }) }
                   \biggr\} 
%******************************************************
 \nn[3mm] & + & 
              \scat{1\to3}(\ana, \Ana,\ana)
              \,\gt^2\mZ^2\,\biggl\{ 
% \nn & & + 
                     - \frac{ 
                       \s{13}^{ }\, \Proj \cdot \P^{ }_2 
                     }{
                 ({ \Sq{\s{12}^{ }}{\mZ} }) ({ \Sq{\s{23}^{ }}{\mZ} }) }
                   \biggr\}
%******************************************************
 \nn[3mm] & + & 
              \scat{1\to3}(\ana, \Aea,\aea)
              \, 4 \gt^2 \cWW^{ } \mW^2 
                  \biggl\{ 
% \nn & & 
                   - \frac{
                       \s{13}^{ }\, \Proj \cdot\P^{ }_2
                     }{
                   ({ \Sq{\s{12}^{ }}{\mW} }) ({ \Sq{\s{23}^{ }}{\mZ} }) }
                   \biggr\} 
%******************************************************
 \nn[3mm] & + & 
              \sum_b\,\scat{1\to3}(\ana, \An,\an)
                   \,\gt^2\mZ^2\,\biggl\{  
% \nn & & + 
                    - \frac{ 
                       \s{13}^{ }\, \Proj \cdot\P^{ }_2 
                     }{ ({ \Sq{\s{23}^{ }}{\mZ} })^2 }
                   \biggr\} 
%******************************************************
 \nn[3mm] & + & 
              \sum_b\,\scat{1\to3}(\ana, \Ae,\aeb)\,\gt^2\mZ^2\,\biggl\{ 
% \nn & & +  
                    - \frac{ 
                       \cWW^2 \s{13}^{ } \, \Proj\cdot\P^{ }_2
                      +
                        4 \sW^4 \s{12}^{ } \, \Proj \cdot\P^{ }_3
                    }{ ({ \Sq{\s{23}^{ }}{\mZ} })^2 }
                   \biggr\} 
%******************************************************
 \nn[3mm] & + & 
              \sum_b\,\scat{1\to3}(\aea, \Ae,\an)\,
                   4 g_2^2\mW^2\, \biggl\{ 
% \nn & & +  
                    - \frac{ 
                      \s{13}^{ }\, \Proj \cdot\P^{ }_2   
                     }{ ({ \Sq{\s{23}^{ }}{\mW} })^2 }
                   \biggr\} 
 \;. \la{leptonic_broken}
\ea}\normalsize

%%%%%%%%%%%%%%%%%%%%%%%% BIBLIO %%%%%%%%%%%%%%%%%%%%%%%%%%%%%%%%%%%%%%%%%%
%

\small{
 
}

\end{document}